\documentclass[longauth]{aaEC}

\usepackage{graphicx}
\usepackage{natbib}
\usepackage{scalerel}
\usepackage{lastpage}

\newcommand{\sigmav}{\sigma_{v}}

\usepackage[table]{xcolor}

\usepackage{txfonts}
\usepackage[pdfencoding=auto,psdextra]{hyperref}
\hypersetup{
    colorlinks=true,
    linkcolor=blue,
    filecolor=magenta,      
    urlcolor=blue,
    citecolor=blue
}
\makeatletter
\renewcommand*\aa@pageof{, page \thepage{} of \pageref*{LastPage}}
\makeatother

\usepackage[utf8]{inputenc}

\usepackage[switch, modulo]{lineno}

\usepackage{euclid}

\begin{document}
%
%

\title{Euclid Quick Data Release (Q1)} \subtitle{The Strong Lensing Discovery Engine B -- Early strong lens candidates from visual inspection of high velocity dispersion galaxies}

   

\newcommand{\orcid}[1]{} 
\author{Euclid Collaboration: K.~Rojas\orcid{0000-0003-1391-6854}\thanks{\email{karina.rojas@fhnw.ch}}\inst{\ref{aff1},\ref{aff2}}
\and T.~E.~Collett\orcid{0000-0001-5564-3140}\inst{\ref{aff2}}
\and J.~A.~Acevedo~Barroso\orcid{0000-0002-9654-1711}\inst{\ref{aff3}}
\and J.~W.~Nightingale\orcid{0000-0002-8987-7401}\inst{\ref{aff4}}
\and D.~Stern\orcid{0000-0003-2686-9241}\inst{\ref{aff5}}
\and L.~A.~Moustakas\orcid{0000-0003-3030-2360}\inst{\ref{aff5}}
\and S.~Schuldt\orcid{0000-0003-2497-6334}\inst{\ref{aff6},\ref{aff7}}
\and G.~Despali\orcid{0000-0001-6150-4112}\inst{\ref{aff8},\ref{aff9},\ref{aff10}}
\and A.~Melo\orcid{0000-0002-6449-3970}\inst{\ref{aff11},\ref{aff12}}
\and M.~Walmsley\orcid{0000-0002-6408-4181}\inst{\ref{aff13},\ref{aff14}}
\and D.~J.~Ballard\orcid{0009-0003-3198-7151}\inst{\ref{aff2},\ref{aff15}}
\and W.~J.~R.~Enzi\orcid{0009-0004-2992-3148}\inst{\ref{aff2}}
\and T.~Li\orcid{0009-0005-5008-0381}\inst{\ref{aff2}}
\and A.~Sainz~de~Murieta\inst{\ref{aff2}}
\and I.~T.~Andika\orcid{0000-0001-6102-9526}\inst{\ref{aff12},\ref{aff11}}
\and B.~Cl\'ement\orcid{0000-0002-7966-3661}\inst{\ref{aff3},\ref{aff16}}
\and F.~Courbin\orcid{0000-0003-0758-6510}\inst{\ref{aff17},\ref{aff18}}
\and L.~R.~Ecker\orcid{0009-0005-3508-2469}\inst{\ref{aff19},\ref{aff20}}
\and R.~Gavazzi\orcid{0000-0002-5540-6935}\inst{\ref{aff21},\ref{aff22}}
\and N.~Jackson\inst{\ref{aff14}}
\and A.~Kov\'acs\orcid{0000-0002-5825-579X}\inst{\ref{aff23},\ref{aff24}}
\and P.~Matavulj\orcid{0000-0003-0229-7189}\inst{\ref{aff1}}
\and M.~Meneghetti\orcid{0000-0003-1225-7084}\inst{\ref{aff9},\ref{aff10}}
\and S.~Serjeant\orcid{0000-0002-0517-7943}\inst{\ref{aff25}}
\and D.~Sluse\orcid{0000-0001-6116-2095}\inst{\ref{aff26}}
\and C.~Tortora\orcid{0000-0001-7958-6531}\inst{\ref{aff27}}
\and A.~Verma\orcid{0000-0002-0730-0781}\inst{\ref{aff28}}
\and L.~Marchetti\orcid{0000-0003-3948-7621}\inst{\ref{aff29},\ref{aff30},\ref{aff31}}
\and C.~M.~O'Riordan\orcid{0000-0003-2227-1998}\inst{\ref{aff11}}
\and K.~McCarthy\orcid{0000-0001-6857-018X}\inst{\ref{aff5}}
\and S.~H.~Suyu\orcid{0000-0001-5568-6052}\inst{\ref{aff12},\ref{aff11}}
\and R.~B.~Metcalf\orcid{0000-0003-3167-2574}\inst{\ref{aff8},\ref{aff9}}
\and N.~Aghanim\orcid{0000-0002-6688-8992}\inst{\ref{aff32}}
\and B.~Altieri\orcid{0000-0003-3936-0284}\inst{\ref{aff33}}
\and A.~Amara\inst{\ref{aff34}}
\and S.~Andreon\orcid{0000-0002-2041-8784}\inst{\ref{aff35}}
\and N.~Auricchio\orcid{0000-0003-4444-8651}\inst{\ref{aff9}}
\and H.~Aussel\orcid{0000-0002-1371-5705}\inst{\ref{aff36}}
\and C.~Baccigalupi\orcid{0000-0002-8211-1630}\inst{\ref{aff37},\ref{aff38},\ref{aff39},\ref{aff40}}
\and M.~Baldi\orcid{0000-0003-4145-1943}\inst{\ref{aff41},\ref{aff9},\ref{aff10}}
\and A.~Balestra\orcid{0000-0002-6967-261X}\inst{\ref{aff42}}
\and S.~Bardelli\orcid{0000-0002-8900-0298}\inst{\ref{aff9}}
\and P.~Battaglia\orcid{0000-0002-7337-5909}\inst{\ref{aff9}}
\and R.~Bender\orcid{0000-0001-7179-0626}\inst{\ref{aff20},\ref{aff19}}
\and A.~Biviano\orcid{0000-0002-0857-0732}\inst{\ref{aff38},\ref{aff37}}
\and A.~Bonchi\orcid{0000-0002-2667-5482}\inst{\ref{aff43}}
\and E.~Branchini\orcid{0000-0002-0808-6908}\inst{\ref{aff44},\ref{aff45},\ref{aff35}}
\and M.~Brescia\orcid{0000-0001-9506-5680}\inst{\ref{aff46},\ref{aff27}}
\and J.~Brinchmann\orcid{0000-0003-4359-8797}\inst{\ref{aff47},\ref{aff48}}
\and S.~Camera\orcid{0000-0003-3399-3574}\inst{\ref{aff49},\ref{aff50},\ref{aff51}}
\and G.~Ca\~nas-Herrera\orcid{0000-0003-2796-2149}\inst{\ref{aff52},\ref{aff53},\ref{aff54}}
\and V.~Capobianco\orcid{0000-0002-3309-7692}\inst{\ref{aff51}}
\and C.~Carbone\orcid{0000-0003-0125-3563}\inst{\ref{aff7}}
\and V.~F.~Cardone\inst{\ref{aff55},\ref{aff56}}
\and J.~Carretero\orcid{0000-0002-3130-0204}\inst{\ref{aff57},\ref{aff58}}
\and S.~Casas\orcid{0000-0002-4751-5138}\inst{\ref{aff59}}
\and M.~Castellano\orcid{0000-0001-9875-8263}\inst{\ref{aff55}}
\and G.~Castignani\orcid{0000-0001-6831-0687}\inst{\ref{aff9}}
\and S.~Cavuoti\orcid{0000-0002-3787-4196}\inst{\ref{aff27},\ref{aff60}}
\and K.~C.~Chambers\orcid{0000-0001-6965-7789}\inst{\ref{aff61}}
\and A.~Cimatti\inst{\ref{aff62}}
\and C.~Colodro-Conde\inst{\ref{aff63}}
\and G.~Congedo\orcid{0000-0003-2508-0046}\inst{\ref{aff64}}
\and C.~J.~Conselice\orcid{0000-0003-1949-7638}\inst{\ref{aff14}}
\and L.~Conversi\orcid{0000-0002-6710-8476}\inst{\ref{aff65},\ref{aff33}}
\and Y.~Copin\orcid{0000-0002-5317-7518}\inst{\ref{aff66}}
\and H.~M.~Courtois\orcid{0000-0003-0509-1776}\inst{\ref{aff67}}
\and M.~Cropper\orcid{0000-0003-4571-9468}\inst{\ref{aff68}}
\and A.~Da~Silva\orcid{0000-0002-6385-1609}\inst{\ref{aff69},\ref{aff70}}
\and H.~Degaudenzi\orcid{0000-0002-5887-6799}\inst{\ref{aff71}}
\and G.~De~Lucia\orcid{0000-0002-6220-9104}\inst{\ref{aff38}}
\and A.~M.~Di~Giorgio\orcid{0000-0002-4767-2360}\inst{\ref{aff72}}
\and C.~Dolding\orcid{0009-0003-7199-6108}\inst{\ref{aff68}}
\and H.~Dole\orcid{0000-0002-9767-3839}\inst{\ref{aff32}}
\and F.~Dubath\orcid{0000-0002-6533-2810}\inst{\ref{aff71}}
\and X.~Dupac\inst{\ref{aff33}}
\and S.~Escoffier\orcid{0000-0002-2847-7498}\inst{\ref{aff73}}
\and M.~Fabricius\orcid{0000-0002-7025-6058}\inst{\ref{aff20},\ref{aff19}}
\and M.~Farina\orcid{0000-0002-3089-7846}\inst{\ref{aff72}}
\and R.~Farinelli\inst{\ref{aff9}}
\and F.~Faustini\orcid{0000-0001-6274-5145}\inst{\ref{aff43},\ref{aff55}}
\and S.~Ferriol\inst{\ref{aff66}}
\and F.~Finelli\orcid{0000-0002-6694-3269}\inst{\ref{aff9},\ref{aff74}}
\and S.~Fotopoulou\orcid{0000-0002-9686-254X}\inst{\ref{aff75}}
\and M.~Frailis\orcid{0000-0002-7400-2135}\inst{\ref{aff38}}
\and E.~Franceschi\orcid{0000-0002-0585-6591}\inst{\ref{aff9}}
\and S.~Galeotta\orcid{0000-0002-3748-5115}\inst{\ref{aff38}}
\and K.~George\orcid{0000-0002-1734-8455}\inst{\ref{aff19}}
\and W.~Gillard\orcid{0000-0003-4744-9748}\inst{\ref{aff73}}
\and B.~Gillis\orcid{0000-0002-4478-1270}\inst{\ref{aff64}}
\and C.~Giocoli\orcid{0000-0002-9590-7961}\inst{\ref{aff9},\ref{aff10}}
\and P.~G\'omez-Alvarez\orcid{0000-0002-8594-5358}\inst{\ref{aff76},\ref{aff33}}
\and J.~Gracia-Carpio\inst{\ref{aff20}}
\and B.~R.~Granett\orcid{0000-0003-2694-9284}\inst{\ref{aff35}}
\and A.~Grazian\orcid{0000-0002-5688-0663}\inst{\ref{aff42}}
\and F.~Grupp\inst{\ref{aff20},\ref{aff19}}
\and L.~Guzzo\orcid{0000-0001-8264-5192}\inst{\ref{aff6},\ref{aff35},\ref{aff77}}
\and S.~Gwyn\orcid{0000-0001-8221-8406}\inst{\ref{aff78}}
\and S.~V.~H.~Haugan\orcid{0000-0001-9648-7260}\inst{\ref{aff79}}
\and W.~Holmes\inst{\ref{aff5}}
\and I.~M.~Hook\orcid{0000-0002-2960-978X}\inst{\ref{aff80}}
\and F.~Hormuth\inst{\ref{aff81}}
\and A.~Hornstrup\orcid{0000-0002-3363-0936}\inst{\ref{aff82},\ref{aff83}}
\and P.~Hudelot\inst{\ref{aff22}}
\and K.~Jahnke\orcid{0000-0003-3804-2137}\inst{\ref{aff84}}
\and M.~Jhabvala\inst{\ref{aff85}}
\and E.~Keih\"anen\orcid{0000-0003-1804-7715}\inst{\ref{aff86}}
\and S.~Kermiche\orcid{0000-0002-0302-5735}\inst{\ref{aff73}}
\and A.~Kiessling\orcid{0000-0002-2590-1273}\inst{\ref{aff5}}
\and B.~Kubik\orcid{0009-0006-5823-4880}\inst{\ref{aff66}}
\and K.~Kuijken\orcid{0000-0002-3827-0175}\inst{\ref{aff54}}
\and M.~K\"ummel\orcid{0000-0003-2791-2117}\inst{\ref{aff19}}
\and M.~Kunz\orcid{0000-0002-3052-7394}\inst{\ref{aff87}}
\and H.~Kurki-Suonio\orcid{0000-0002-4618-3063}\inst{\ref{aff88},\ref{aff89}}
\and Q.~Le~Boulc'h\inst{\ref{aff90}}
\and A.~M.~C.~Le~Brun\orcid{0000-0002-0936-4594}\inst{\ref{aff91}}
\and D.~Le~Mignant\orcid{0000-0002-5339-5515}\inst{\ref{aff21}}
\and P.~Liebing\inst{\ref{aff68}}
\and S.~Ligori\orcid{0000-0003-4172-4606}\inst{\ref{aff51}}
\and P.~B.~Lilje\orcid{0000-0003-4324-7794}\inst{\ref{aff79}}
\and V.~Lindholm\orcid{0000-0003-2317-5471}\inst{\ref{aff88},\ref{aff89}}
\and I.~Lloro\orcid{0000-0001-5966-1434}\inst{\ref{aff92}}
\and G.~Mainetti\orcid{0000-0003-2384-2377}\inst{\ref{aff90}}
\and D.~Maino\inst{\ref{aff6},\ref{aff7},\ref{aff77}}
\and E.~Maiorano\orcid{0000-0003-2593-4355}\inst{\ref{aff9}}
\and O.~Mansutti\orcid{0000-0001-5758-4658}\inst{\ref{aff38}}
\and S.~Marcin\inst{\ref{aff93}}
\and O.~Marggraf\orcid{0000-0001-7242-3852}\inst{\ref{aff94}}
\and M.~Martinelli\orcid{0000-0002-6943-7732}\inst{\ref{aff55},\ref{aff56}}
\and N.~Martinet\orcid{0000-0003-2786-7790}\inst{\ref{aff21}}
\and F.~Marulli\orcid{0000-0002-8850-0303}\inst{\ref{aff8},\ref{aff9},\ref{aff10}}
\and R.~Massey\orcid{0000-0002-6085-3780}\inst{\ref{aff95}}
\and S.~Maurogordato\inst{\ref{aff96}}
\and H.~J.~McCracken\orcid{0000-0002-9489-7765}\inst{\ref{aff22}}
\and E.~Medinaceli\orcid{0000-0002-4040-7783}\inst{\ref{aff9}}
\and S.~Mei\orcid{0000-0002-2849-559X}\inst{\ref{aff97},\ref{aff98}}
\and M.~Melchior\inst{\ref{aff1}}
\and Y.~Mellier\inst{\ref{aff99},\ref{aff22}}
\and E.~Merlin\orcid{0000-0001-6870-8900}\inst{\ref{aff55}}
\and G.~Meylan\inst{\ref{aff3}}
\and A.~Mora\orcid{0000-0002-1922-8529}\inst{\ref{aff100}}
\and M.~Moresco\orcid{0000-0002-7616-7136}\inst{\ref{aff8},\ref{aff9}}
\and L.~Moscardini\orcid{0000-0002-3473-6716}\inst{\ref{aff8},\ref{aff9},\ref{aff10}}
\and R.~Nakajima\orcid{0009-0009-1213-7040}\inst{\ref{aff94}}
\and C.~Neissner\orcid{0000-0001-8524-4968}\inst{\ref{aff101},\ref{aff58}}
\and R.~C.~Nichol\orcid{0000-0003-0939-6518}\inst{\ref{aff34}}
\and S.-M.~Niemi\inst{\ref{aff52}}
\and C.~Padilla\orcid{0000-0001-7951-0166}\inst{\ref{aff101}}
\and S.~Paltani\orcid{0000-0002-8108-9179}\inst{\ref{aff71}}
\and F.~Pasian\orcid{0000-0002-4869-3227}\inst{\ref{aff38}}
\and K.~Pedersen\inst{\ref{aff102}}
\and W.~J.~Percival\orcid{0000-0002-0644-5727}\inst{\ref{aff103},\ref{aff104},\ref{aff105}}
\and V.~Pettorino\inst{\ref{aff52}}
\and S.~Pires\orcid{0000-0002-0249-2104}\inst{\ref{aff36}}
\and G.~Polenta\orcid{0000-0003-4067-9196}\inst{\ref{aff43}}
\and M.~Poncet\inst{\ref{aff106}}
\and L.~A.~Popa\inst{\ref{aff107}}
\and L.~Pozzetti\orcid{0000-0001-7085-0412}\inst{\ref{aff9}}
\and F.~Raison\orcid{0000-0002-7819-6918}\inst{\ref{aff20}}
\and R.~Rebolo\orcid{0000-0003-3767-7085}\inst{\ref{aff63},\ref{aff108},\ref{aff109}}
\and A.~Renzi\orcid{0000-0001-9856-1970}\inst{\ref{aff110},\ref{aff111}}
\and J.~Rhodes\orcid{0000-0002-4485-8549}\inst{\ref{aff5}}
\and G.~Riccio\inst{\ref{aff27}}
\and E.~Romelli\orcid{0000-0003-3069-9222}\inst{\ref{aff38}}
\and M.~Roncarelli\orcid{0000-0001-9587-7822}\inst{\ref{aff9}}
\and R.~Saglia\orcid{0000-0003-0378-7032}\inst{\ref{aff19},\ref{aff20}}
\and Z.~Sakr\orcid{0000-0002-4823-3757}\inst{\ref{aff112},\ref{aff113},\ref{aff114}}
\and A.~G.~S\'anchez\orcid{0000-0003-1198-831X}\inst{\ref{aff20}}
\and D.~Sapone\orcid{0000-0001-7089-4503}\inst{\ref{aff115}}
\and B.~Sartoris\orcid{0000-0003-1337-5269}\inst{\ref{aff19},\ref{aff38}}
\and J.~A.~Schewtschenko\orcid{0000-0002-4913-6393}\inst{\ref{aff64}}
\and M.~Schirmer\orcid{0000-0003-2568-9994}\inst{\ref{aff84}}
\and P.~Schneider\orcid{0000-0001-8561-2679}\inst{\ref{aff94}}
\and T.~Schrabback\orcid{0000-0002-6987-7834}\inst{\ref{aff116}}
\and A.~Secroun\orcid{0000-0003-0505-3710}\inst{\ref{aff73}}
\and G.~Seidel\orcid{0000-0003-2907-353X}\inst{\ref{aff84}}
\and M.~Seiffert\orcid{0000-0002-7536-9393}\inst{\ref{aff5}}
\and S.~Serrano\orcid{0000-0002-0211-2861}\inst{\ref{aff117},\ref{aff118},\ref{aff119}}
\and P.~Simon\inst{\ref{aff94}}
\and C.~Sirignano\orcid{0000-0002-0995-7146}\inst{\ref{aff110},\ref{aff111}}
\and G.~Sirri\orcid{0000-0003-2626-2853}\inst{\ref{aff10}}
\and L.~Stanco\orcid{0000-0002-9706-5104}\inst{\ref{aff111}}
\and J.~Steinwagner\orcid{0000-0001-7443-1047}\inst{\ref{aff20}}
\and P.~Tallada-Cresp\'{i}\orcid{0000-0002-1336-8328}\inst{\ref{aff57},\ref{aff58}}
\and A.~N.~Taylor\inst{\ref{aff64}}
\and I.~Tereno\inst{\ref{aff69},\ref{aff120}}
\and S.~Toft\orcid{0000-0003-3631-7176}\inst{\ref{aff121},\ref{aff122}}
\and R.~Toledo-Moreo\orcid{0000-0002-2997-4859}\inst{\ref{aff123}}
\and F.~Torradeflot\orcid{0000-0003-1160-1517}\inst{\ref{aff58},\ref{aff57}}
\and I.~Tutusaus\orcid{0000-0002-3199-0399}\inst{\ref{aff113}}
\and L.~Valenziano\orcid{0000-0002-1170-0104}\inst{\ref{aff9},\ref{aff74}}
\and J.~Valiviita\orcid{0000-0001-6225-3693}\inst{\ref{aff88},\ref{aff89}}
\and T.~Vassallo\orcid{0000-0001-6512-6358}\inst{\ref{aff19},\ref{aff38}}
\and G.~Verdoes~Kleijn\orcid{0000-0001-5803-2580}\inst{\ref{aff124}}
\and A.~Veropalumbo\orcid{0000-0003-2387-1194}\inst{\ref{aff35},\ref{aff45},\ref{aff44}}
\and Y.~Wang\orcid{0000-0002-4749-2984}\inst{\ref{aff125}}
\and J.~Weller\orcid{0000-0002-8282-2010}\inst{\ref{aff19},\ref{aff20}}
\and A.~Zacchei\orcid{0000-0003-0396-1192}\inst{\ref{aff38},\ref{aff37}}
\and G.~Zamorani\orcid{0000-0002-2318-301X}\inst{\ref{aff9}}
\and F.~M.~Zerbi\inst{\ref{aff35}}
\and E.~Zucca\orcid{0000-0002-5845-8132}\inst{\ref{aff9}}
\and M.~Ballardini\orcid{0000-0003-4481-3559}\inst{\ref{aff126},\ref{aff127},\ref{aff9}}
\and M.~Bolzonella\orcid{0000-0003-3278-4607}\inst{\ref{aff9}}
\and E.~Bozzo\orcid{0000-0002-8201-1525}\inst{\ref{aff71}}
\and C.~Burigana\orcid{0000-0002-3005-5796}\inst{\ref{aff31},\ref{aff74}}
\and R.~Cabanac\orcid{0000-0001-6679-2600}\inst{\ref{aff113}}
\and A.~Cappi\inst{\ref{aff9},\ref{aff96}}
\and D.~Di~Ferdinando\inst{\ref{aff10}}
\and J.~A.~Escartin~Vigo\inst{\ref{aff20}}
\and L.~Gabarra\orcid{0000-0002-8486-8856}\inst{\ref{aff28}}
\and J.~Mart\'{i}n-Fleitas\orcid{0000-0002-8594-569X}\inst{\ref{aff100}}
\and S.~Matthew\orcid{0000-0001-8448-1697}\inst{\ref{aff64}}
\and N.~Mauri\orcid{0000-0001-8196-1548}\inst{\ref{aff62},\ref{aff10}}
\and A.~Pezzotta\orcid{0000-0003-0726-2268}\inst{\ref{aff128},\ref{aff20}}
\and M.~P\"ontinen\orcid{0000-0001-5442-2530}\inst{\ref{aff88}}
\and C.~Porciani\orcid{0000-0002-7797-2508}\inst{\ref{aff94}}
\and I.~Risso\orcid{0000-0003-2525-7761}\inst{\ref{aff129}}
\and V.~Scottez\inst{\ref{aff99},\ref{aff130}}
\and M.~Sereno\orcid{0000-0003-0302-0325}\inst{\ref{aff9},\ref{aff10}}
\and M.~Tenti\orcid{0000-0002-4254-5901}\inst{\ref{aff10}}
\and M.~Viel\orcid{0000-0002-2642-5707}\inst{\ref{aff37},\ref{aff38},\ref{aff40},\ref{aff39},\ref{aff131}}
\and M.~Wiesmann\orcid{0009-0000-8199-5860}\inst{\ref{aff79}}
\and Y.~Akrami\orcid{0000-0002-2407-7956}\inst{\ref{aff132},\ref{aff133}}
\and S.~Alvi\orcid{0000-0001-5779-8568}\inst{\ref{aff126}}
\and S.~Anselmi\orcid{0000-0002-3579-9583}\inst{\ref{aff111},\ref{aff110},\ref{aff134}}
\and M.~Archidiacono\orcid{0000-0003-4952-9012}\inst{\ref{aff6},\ref{aff77}}
\and F.~Atrio-Barandela\orcid{0000-0002-2130-2513}\inst{\ref{aff135}}
\and C.~Benoist\inst{\ref{aff96}}
\and K.~Benson\inst{\ref{aff68}}
\and P.~Bergamini\orcid{0000-0003-1383-9414}\inst{\ref{aff6},\ref{aff9}}
\and D.~Bertacca\orcid{0000-0002-2490-7139}\inst{\ref{aff110},\ref{aff42},\ref{aff111}}
\and M.~Bethermin\orcid{0000-0002-3915-2015}\inst{\ref{aff136}}
\and A.~Blanchard\orcid{0000-0001-8555-9003}\inst{\ref{aff113}}
\and L.~Blot\orcid{0000-0002-9622-7167}\inst{\ref{aff137},\ref{aff134}}
\and M.~L.~Brown\orcid{0000-0002-0370-8077}\inst{\ref{aff14}}
\and S.~Bruton\orcid{0000-0002-6503-5218}\inst{\ref{aff138}}
\and A.~Calabro\orcid{0000-0003-2536-1614}\inst{\ref{aff55}}
\and B.~Camacho~Quevedo\orcid{0000-0002-8789-4232}\inst{\ref{aff117},\ref{aff119}}
\and F.~Caro\inst{\ref{aff55}}
\and C.~S.~Carvalho\inst{\ref{aff120}}
\and T.~Castro\orcid{0000-0002-6292-3228}\inst{\ref{aff38},\ref{aff39},\ref{aff37},\ref{aff131}}
\and F.~Cogato\orcid{0000-0003-4632-6113}\inst{\ref{aff8},\ref{aff9}}
\and A.~R.~Cooray\orcid{0000-0002-3892-0190}\inst{\ref{aff139}}
\and O.~Cucciati\orcid{0000-0002-9336-7551}\inst{\ref{aff9}}
\and S.~Davini\orcid{0000-0003-3269-1718}\inst{\ref{aff45}}
\and F.~De~Paolis\orcid{0000-0001-6460-7563}\inst{\ref{aff140},\ref{aff141},\ref{aff142}}
\and G.~Desprez\orcid{0000-0001-8325-1742}\inst{\ref{aff124}}
\and A.~D\'iaz-S\'anchez\orcid{0000-0003-0748-4768}\inst{\ref{aff143}}
\and J.~J.~Diaz\inst{\ref{aff63}}
\and S.~Di~Domizio\orcid{0000-0003-2863-5895}\inst{\ref{aff44},\ref{aff45}}
\and J.~M.~Diego\orcid{0000-0001-9065-3926}\inst{\ref{aff144}}
\and P.-A.~Duc\orcid{0000-0003-3343-6284}\inst{\ref{aff136}}
\and A.~Enia\orcid{0000-0002-0200-2857}\inst{\ref{aff41},\ref{aff9}}
\and Y.~Fang\inst{\ref{aff19}}
\and A.~G.~Ferrari\orcid{0009-0005-5266-4110}\inst{\ref{aff10}}
\and P.~G.~Ferreira\orcid{0000-0002-3021-2851}\inst{\ref{aff28}}
\and A.~Finoguenov\orcid{0000-0002-4606-5403}\inst{\ref{aff88}}
\and A.~Fontana\orcid{0000-0003-3820-2823}\inst{\ref{aff55}}
\and A.~Franco\orcid{0000-0002-4761-366X}\inst{\ref{aff141},\ref{aff140},\ref{aff142}}
\and K.~Ganga\orcid{0000-0001-8159-8208}\inst{\ref{aff97}}
\and J.~Garc\'ia-Bellido\orcid{0000-0002-9370-8360}\inst{\ref{aff132}}
\and T.~Gasparetto\orcid{0000-0002-7913-4866}\inst{\ref{aff38}}
\and V.~Gautard\inst{\ref{aff145}}
\and E.~Gaztanaga\orcid{0000-0001-9632-0815}\inst{\ref{aff119},\ref{aff117},\ref{aff2}}
\and F.~Giacomini\orcid{0000-0002-3129-2814}\inst{\ref{aff10}}
\and F.~Gianotti\orcid{0000-0003-4666-119X}\inst{\ref{aff9}}
\and G.~Gozaliasl\orcid{0000-0002-0236-919X}\inst{\ref{aff146},\ref{aff88}}
\and M.~Guidi\orcid{0000-0001-9408-1101}\inst{\ref{aff41},\ref{aff9}}
\and C.~M.~Gutierrez\orcid{0000-0001-7854-783X}\inst{\ref{aff147}}
\and A.~Hall\orcid{0000-0002-3139-8651}\inst{\ref{aff64}}
\and W.~G.~Hartley\inst{\ref{aff71}}
\and C.~Hern\'andez-Monteagudo\orcid{0000-0001-5471-9166}\inst{\ref{aff109},\ref{aff63}}
\and H.~Hildebrandt\orcid{0000-0002-9814-3338}\inst{\ref{aff148}}
\and J.~Hjorth\orcid{0000-0002-4571-2306}\inst{\ref{aff102}}
\and J.~J.~E.~Kajava\orcid{0000-0002-3010-8333}\inst{\ref{aff149},\ref{aff150}}
\and Y.~Kang\orcid{0009-0000-8588-7250}\inst{\ref{aff71}}
\and V.~Kansal\orcid{0000-0002-4008-6078}\inst{\ref{aff151},\ref{aff152}}
\and D.~Karagiannis\orcid{0000-0002-4927-0816}\inst{\ref{aff126},\ref{aff153}}
\and K.~Kiiveri\inst{\ref{aff86}}
\and C.~C.~Kirkpatrick\inst{\ref{aff86}}
\and S.~Kruk\orcid{0000-0001-8010-8879}\inst{\ref{aff33}}
\and J.~Le~Graet\orcid{0000-0001-6523-7971}\inst{\ref{aff73}}
\and L.~Legrand\orcid{0000-0003-0610-5252}\inst{\ref{aff154},\ref{aff155}}
\and M.~Lembo\orcid{0000-0002-5271-5070}\inst{\ref{aff126},\ref{aff127}}
\and F.~Lepori\orcid{0009-0000-5061-7138}\inst{\ref{aff156}}
\and G.~Leroy\orcid{0009-0004-2523-4425}\inst{\ref{aff157},\ref{aff95}}
\and G.~F.~Lesci\orcid{0000-0002-4607-2830}\inst{\ref{aff8},\ref{aff9}}
\and J.~Lesgourgues\orcid{0000-0001-7627-353X}\inst{\ref{aff59}}
\and L.~Leuzzi\orcid{0009-0006-4479-7017}\inst{\ref{aff8},\ref{aff9}}
\and T.~I.~Liaudat\orcid{0000-0002-9104-314X}\inst{\ref{aff158}}
\and A.~Loureiro\orcid{0000-0002-4371-0876}\inst{\ref{aff159},\ref{aff160}}
\and J.~Macias-Perez\orcid{0000-0002-5385-2763}\inst{\ref{aff161}}
\and G.~Maggio\orcid{0000-0003-4020-4836}\inst{\ref{aff38}}
\and M.~Magliocchetti\orcid{0000-0001-9158-4838}\inst{\ref{aff72}}
\and E.~A.~Magnier\orcid{0000-0002-7965-2815}\inst{\ref{aff61}}
\and F.~Mannucci\orcid{0000-0002-4803-2381}\inst{\ref{aff162}}
\and R.~Maoli\orcid{0000-0002-6065-3025}\inst{\ref{aff163},\ref{aff55}}
\and C.~J.~A.~P.~Martins\orcid{0000-0002-4886-9261}\inst{\ref{aff164},\ref{aff47}}
\and L.~Maurin\orcid{0000-0002-8406-0857}\inst{\ref{aff32}}
\and M.~Miluzio\inst{\ref{aff33},\ref{aff165}}
\and P.~Monaco\orcid{0000-0003-2083-7564}\inst{\ref{aff166},\ref{aff38},\ref{aff39},\ref{aff37}}
\and C.~Moretti\orcid{0000-0003-3314-8936}\inst{\ref{aff40},\ref{aff131},\ref{aff38},\ref{aff37},\ref{aff39}}
\and G.~Morgante\inst{\ref{aff9}}
\and S.~Nadathur\orcid{0000-0001-9070-3102}\inst{\ref{aff2}}
\and K.~Naidoo\orcid{0000-0002-9182-1802}\inst{\ref{aff2}}
\and A.~Navarro-Alsina\orcid{0000-0002-3173-2592}\inst{\ref{aff94}}
\and S.~Nesseris\orcid{0000-0002-0567-0324}\inst{\ref{aff132}}
\and F.~Passalacqua\orcid{0000-0002-8606-4093}\inst{\ref{aff110},\ref{aff111}}
\and K.~Paterson\orcid{0000-0001-8340-3486}\inst{\ref{aff84}}
\and L.~Patrizii\inst{\ref{aff10}}
\and A.~Pisani\orcid{0000-0002-6146-4437}\inst{\ref{aff73},\ref{aff167}}
\and D.~Potter\orcid{0000-0002-0757-5195}\inst{\ref{aff156}}
\and S.~Quai\orcid{0000-0002-0449-8163}\inst{\ref{aff8},\ref{aff9}}
\and M.~Radovich\orcid{0000-0002-3585-866X}\inst{\ref{aff42}}
\and P.-F.~Rocci\inst{\ref{aff32}}
\and S.~Sacquegna\orcid{0000-0002-8433-6630}\inst{\ref{aff140},\ref{aff141},\ref{aff142}}
\and M.~Sahl\'en\orcid{0000-0003-0973-4804}\inst{\ref{aff168}}
\and D.~B.~Sanders\orcid{0000-0002-1233-9998}\inst{\ref{aff61}}
\and E.~Sarpa\orcid{0000-0002-1256-655X}\inst{\ref{aff40},\ref{aff131},\ref{aff39}}
\and C.~Scarlata\orcid{0000-0002-9136-8876}\inst{\ref{aff169}}
\and A.~Schneider\orcid{0000-0001-7055-8104}\inst{\ref{aff156}}
\and D.~Sciotti\orcid{0009-0008-4519-2620}\inst{\ref{aff55},\ref{aff56}}
\and E.~Sellentin\inst{\ref{aff170},\ref{aff54}}
\and L.~C.~Smith\orcid{0000-0002-3259-2771}\inst{\ref{aff171}}
\and K.~Tanidis\orcid{0000-0001-9843-5130}\inst{\ref{aff28}}
\and G.~Testera\inst{\ref{aff45}}
\and R.~Teyssier\orcid{0000-0001-7689-0933}\inst{\ref{aff167}}
\and A.~Troja\orcid{0000-0003-0239-4595}\inst{\ref{aff110},\ref{aff111}}
\and M.~Tucci\inst{\ref{aff71}}
\and C.~Valieri\inst{\ref{aff10}}
\and A.~Venhola\orcid{0000-0001-6071-4564}\inst{\ref{aff172}}
\and D.~Vergani\orcid{0000-0003-0898-2216}\inst{\ref{aff9}}
\and G.~Vernardos\orcid{0000-0001-8554-7248}\inst{\ref{aff173},\ref{aff174}}
\and G.~Verza\orcid{0000-0002-1886-8348}\inst{\ref{aff175}}
\and P.~Vielzeuf\orcid{0000-0003-2035-9339}\inst{\ref{aff73}}
\and N.~A.~Walton\orcid{0000-0003-3983-8778}\inst{\ref{aff171}}
\and J.~Wilde\orcid{0000-0002-4460-7379}\inst{\ref{aff17}}
\and D.~Scott\orcid{0000-0002-6878-9840}\inst{\ref{aff176}}}
										   
\institute{University of Applied Sciences and Arts of Northwestern Switzerland, School of Engineering, 5210 Windisch, Switzerland\label{aff1}
\and
Institute of Cosmology and Gravitation, University of Portsmouth, Portsmouth PO1 3FX, UK\label{aff2}
\and
Institute of Physics, Laboratory of Astrophysics, Ecole Polytechnique F\'ed\'erale de Lausanne (EPFL), Observatoire de Sauverny, 1290 Versoix, Switzerland\label{aff3}
\and
School of Mathematics, Statistics and Physics, Newcastle University, Herschel Building, Newcastle-upon-Tyne, NE1 7RU, UK\label{aff4}
\and
Jet Propulsion Laboratory, California Institute of Technology, 4800 Oak Grove Drive, Pasadena, CA, 91109, USA\label{aff5}
\and
Dipartimento di Fisica "Aldo Pontremoli", Universit\`a degli Studi di Milano, Via Celoria 16, 20133 Milano, Italy\label{aff6}
\and
INAF-IASF Milano, Via Alfonso Corti 12, 20133 Milano, Italy\label{aff7}
\and
Dipartimento di Fisica e Astronomia "Augusto Righi" - Alma Mater Studiorum Universit\`a di Bologna, via Piero Gobetti 93/2, 40129 Bologna, Italy\label{aff8}
\and
INAF-Osservatorio di Astrofisica e Scienza dello Spazio di Bologna, Via Piero Gobetti 93/3, 40129 Bologna, Italy\label{aff9}
\and
INFN-Sezione di Bologna, Viale Berti Pichat 6/2, 40127 Bologna, Italy\label{aff10}
\and
Max-Planck-Institut f\"ur Astrophysik, Karl-Schwarzschild-Str.~1, 85748 Garching, Germany\label{aff11}
\and
Technical University of Munich, TUM School of Natural Sciences, Physics Department, James-Franck-Str.~1, 85748 Garching, Germany\label{aff12}
\and
David A. Dunlap Department of Astronomy \& Astrophysics, University of Toronto, 50 St George Street, Toronto, Ontario M5S 3H4, Canada\label{aff13}
\and
Jodrell Bank Centre for Astrophysics, Department of Physics and Astronomy, University of Manchester, Oxford Road, Manchester M13 9PL, UK\label{aff14}
\and
Sydney Institute for Astronomy, School of Physics, University of Sydney, NSW 2006, Australia\label{aff15}
\and
SCITAS, Ecole Polytechnique F\'ed\'erale de Lausanne (EPFL), 1015 Lausanne, Switzerland\label{aff16}
\and
Institut de Ci\`{e}ncies del Cosmos (ICCUB), Universitat de Barcelona (IEEC-UB), Mart\'{i} i Franqu\`{e}s 1, 08028 Barcelona, Spain\label{aff17}
\and
Instituci\'o Catalana de Recerca i Estudis Avan\c{c}ats (ICREA), Passeig de Llu\'{\i}s Companys 23, 08010 Barcelona, Spain\label{aff18}
\and
Universit\"ats-Sternwarte M\"unchen, Fakult\"at f\"ur Physik, Ludwig-Maximilians-Universit\"at M\"unchen, Scheinerstrasse 1, 81679 M\"unchen, Germany\label{aff19}
\and
Max Planck Institute for Extraterrestrial Physics, Giessenbachstr. 1, 85748 Garching, Germany\label{aff20}
\and
Aix-Marseille Universit\'e, CNRS, CNES, LAM, Marseille, France\label{aff21}
\and
Institut d'Astrophysique de Paris, UMR 7095, CNRS, and Sorbonne Universit\'e, 98 bis boulevard Arago, 75014 Paris, France\label{aff22}
\and
MTA-CSFK Lend\"ulet Large-Scale Structure Research Group, Konkoly-Thege Mikl\'os \'ut 15-17, H-1121 Budapest, Hungary\label{aff23}
\and
Konkoly Observatory, HUN-REN CSFK, MTA Centre of Excellence, Budapest, Konkoly Thege Mikl\'os {\'u}t 15-17. H-1121, Hungary\label{aff24}
\and
School of Physical Sciences, The Open University, Milton Keynes, MK7 6AA, UK\label{aff25}
\and
STAR Institute, University of Li{\`e}ge, Quartier Agora, All\'ee du six Ao\^ut 19c, 4000 Li\`ege, Belgium\label{aff26}
\and
INAF-Osservatorio Astronomico di Capodimonte, Via Moiariello 16, 80131 Napoli, Italy\label{aff27}
\and
Department of Physics, Oxford University, Keble Road, Oxford OX1 3RH, UK\label{aff28}
\and
Department of Astronomy, University of Cape Town, Rondebosch, Cape Town, 7700, South Africa\label{aff29}
\and
Inter-University Institute for Data Intensive Astronomy, Department of Astronomy, University of Cape Town, 7701 Rondebosch, Cape Town, South Africa\label{aff30}
\and
INAF, Istituto di Radioastronomia, Via Piero Gobetti 101, 40129 Bologna, Italy\label{aff31}
\and
Universit\'e Paris-Saclay, CNRS, Institut d'astrophysique spatiale, 91405, Orsay, France\label{aff32}
\and
ESAC/ESA, Camino Bajo del Castillo, s/n., Urb. Villafranca del Castillo, 28692 Villanueva de la Ca\~nada, Madrid, Spain\label{aff33}
\and
School of Mathematics and Physics, University of Surrey, Guildford, Surrey, GU2 7XH, UK\label{aff34}
\and
INAF-Osservatorio Astronomico di Brera, Via Brera 28, 20122 Milano, Italy\label{aff35}
\and
Universit\'e Paris-Saclay, Universit\'e Paris Cit\'e, CEA, CNRS, AIM, 91191, Gif-sur-Yvette, France\label{aff36}
\and
IFPU, Institute for Fundamental Physics of the Universe, via Beirut 2, 34151 Trieste, Italy\label{aff37}
\and
INAF-Osservatorio Astronomico di Trieste, Via G. B. Tiepolo 11, 34143 Trieste, Italy\label{aff38}
\and
INFN, Sezione di Trieste, Via Valerio 2, 34127 Trieste TS, Italy\label{aff39}
\and
SISSA, International School for Advanced Studies, Via Bonomea 265, 34136 Trieste TS, Italy\label{aff40}
\and
Dipartimento di Fisica e Astronomia, Universit\`a di Bologna, Via Gobetti 93/2, 40129 Bologna, Italy\label{aff41}
\and
INAF-Osservatorio Astronomico di Padova, Via dell'Osservatorio 5, 35122 Padova, Italy\label{aff42}
\and
Space Science Data Center, Italian Space Agency, via del Politecnico snc, 00133 Roma, Italy\label{aff43}
\and
Dipartimento di Fisica, Universit\`a di Genova, Via Dodecaneso 33, 16146, Genova, Italy\label{aff44}
\and
INFN-Sezione di Genova, Via Dodecaneso 33, 16146, Genova, Italy\label{aff45}
\and
Department of Physics "E. Pancini", University Federico II, Via Cinthia 6, 80126, Napoli, Italy\label{aff46}
\and
Instituto de Astrof\'isica e Ci\^encias do Espa\c{c}o, Universidade do Porto, CAUP, Rua das Estrelas, PT4150-762 Porto, Portugal\label{aff47}
\and
Faculdade de Ci\^encias da Universidade do Porto, Rua do Campo de Alegre, 4150-007 Porto, Portugal\label{aff48}
\and
Dipartimento di Fisica, Universit\`a degli Studi di Torino, Via P. Giuria 1, 10125 Torino, Italy\label{aff49}
\and
INFN-Sezione di Torino, Via P. Giuria 1, 10125 Torino, Italy\label{aff50}
\and
INAF-Osservatorio Astrofisico di Torino, Via Osservatorio 20, 10025 Pino Torinese (TO), Italy\label{aff51}
\and
European Space Agency/ESTEC, Keplerlaan 1, 2201 AZ Noordwijk, The Netherlands\label{aff52}
\and
Institute Lorentz, Leiden University, Niels Bohrweg 2, 2333 CA Leiden, The Netherlands\label{aff53}
\and
Leiden Observatory, Leiden University, Einsteinweg 55, 2333 CC Leiden, The Netherlands\label{aff54}
\and
INAF-Osservatorio Astronomico di Roma, Via Frascati 33, 00078 Monteporzio Catone, Italy\label{aff55}
\and
INFN-Sezione di Roma, Piazzale Aldo Moro, 2 - c/o Dipartimento di Fisica, Edificio G. Marconi, 00185 Roma, Italy\label{aff56}
\and
Centro de Investigaciones Energ\'eticas, Medioambientales y Tecnol\'ogicas (CIEMAT), Avenida Complutense 40, 28040 Madrid, Spain\label{aff57}
\and
Port d'Informaci\'{o} Cient\'{i}fica, Campus UAB, C. Albareda s/n, 08193 Bellaterra (Barcelona), Spain\label{aff58}
\and
Institute for Theoretical Particle Physics and Cosmology (TTK), RWTH Aachen University, 52056 Aachen, Germany\label{aff59}
\and
INFN section of Naples, Via Cinthia 6, 80126, Napoli, Italy\label{aff60}
\and
Institute for Astronomy, University of Hawaii, 2680 Woodlawn Drive, Honolulu, HI 96822, USA\label{aff61}
\and
Dipartimento di Fisica e Astronomia "Augusto Righi" - Alma Mater Studiorum Universit\`a di Bologna, Viale Berti Pichat 6/2, 40127 Bologna, Italy\label{aff62}
\and
Instituto de Astrof\'{\i}sica de Canarias, V\'{\i}a L\'actea, 38205 La Laguna, Tenerife, Spain\label{aff63}
\and
Institute for Astronomy, University of Edinburgh, Royal Observatory, Blackford Hill, Edinburgh EH9 3HJ, UK\label{aff64}
\and
European Space Agency/ESRIN, Largo Galileo Galilei 1, 00044 Frascati, Roma, Italy\label{aff65}
\and
Universit\'e Claude Bernard Lyon 1, CNRS/IN2P3, IP2I Lyon, UMR 5822, Villeurbanne, F-69100, France\label{aff66}
\and
UCB Lyon 1, CNRS/IN2P3, IUF, IP2I Lyon, 4 rue Enrico Fermi, 69622 Villeurbanne, France\label{aff67}
\and
Mullard Space Science Laboratory, University College London, Holmbury St Mary, Dorking, Surrey RH5 6NT, UK\label{aff68}
\and
Departamento de F\'isica, Faculdade de Ci\^encias, Universidade de Lisboa, Edif\'icio C8, Campo Grande, PT1749-016 Lisboa, Portugal\label{aff69}
\and
Instituto de Astrof\'isica e Ci\^encias do Espa\c{c}o, Faculdade de Ci\^encias, Universidade de Lisboa, Campo Grande, 1749-016 Lisboa, Portugal\label{aff70}
\and
Department of Astronomy, University of Geneva, ch. d'Ecogia 16, 1290 Versoix, Switzerland\label{aff71}
\and
INAF-Istituto di Astrofisica e Planetologia Spaziali, via del Fosso del Cavaliere, 100, 00100 Roma, Italy\label{aff72}
\and
Aix-Marseille Universit\'e, CNRS/IN2P3, CPPM, Marseille, France\label{aff73}
\and
INFN-Bologna, Via Irnerio 46, 40126 Bologna, Italy\label{aff74}
\and
School of Physics, HH Wills Physics Laboratory, University of Bristol, Tyndall Avenue, Bristol, BS8 1TL, UK\label{aff75}
\and
FRACTAL S.L.N.E., calle Tulip\'an 2, Portal 13 1A, 28231, Las Rozas de Madrid, Spain\label{aff76}
\and
INFN-Sezione di Milano, Via Celoria 16, 20133 Milano, Italy\label{aff77}
\and
NRC Herzberg, 5071 West Saanich Rd, Victoria, BC V9E 2E7, Canada\label{aff78}
\and
Institute of Theoretical Astrophysics, University of Oslo, P.O. Box 1029 Blindern, 0315 Oslo, Norway\label{aff79}
\and
Department of Physics, Lancaster University, Lancaster, LA1 4YB, UK\label{aff80}
\and
Felix Hormuth Engineering, Goethestr. 17, 69181 Leimen, Germany\label{aff81}
\and
Technical University of Denmark, Elektrovej 327, 2800 Kgs. Lyngby, Denmark\label{aff82}
\and
Cosmic Dawn Center (DAWN), Denmark\label{aff83}
\and
Max-Planck-Institut f\"ur Astronomie, K\"onigstuhl 17, 69117 Heidelberg, Germany\label{aff84}
\and
NASA Goddard Space Flight Center, Greenbelt, MD 20771, USA\label{aff85}
\and
Department of Physics and Helsinki Institute of Physics, Gustaf H\"allstr\"omin katu 2, 00014 University of Helsinki, Finland\label{aff86}
\and
Universit\'e de Gen\`eve, D\'epartement de Physique Th\'eorique and Centre for Astroparticle Physics, 24 quai Ernest-Ansermet, CH-1211 Gen\`eve 4, Switzerland\label{aff87}
\and
Department of Physics, P.O. Box 64, 00014 University of Helsinki, Finland\label{aff88}
\and
Helsinki Institute of Physics, Gustaf H{\"a}llstr{\"o}min katu 2, University of Helsinki, Helsinki, Finland\label{aff89}
\and
Centre de Calcul de l'IN2P3/CNRS, 21 avenue Pierre de Coubertin 69627 Villeurbanne Cedex, France\label{aff90}
\and
Laboratoire d'etude de l'Univers et des phenomenes eXtremes, Observatoire de Paris, Universit\'e PSL, Sorbonne Universit\'e, CNRS, 92190 Meudon, France\label{aff91}
\and
SKA Observatory, Jodrell Bank, Lower Withington, Macclesfield, Cheshire SK11 9FT, UK\label{aff92}
\and
University of Applied Sciences and Arts of Northwestern Switzerland, School of Computer Science, 5210 Windisch, Switzerland\label{aff93}
\and
Universit\"at Bonn, Argelander-Institut f\"ur Astronomie, Auf dem H\"ugel 71, 53121 Bonn, Germany\label{aff94}
\and
Department of Physics, Institute for Computational Cosmology, Durham University, South Road, Durham, DH1 3LE, UK\label{aff95}
\and
Universit\'e C\^{o}te d'Azur, Observatoire de la C\^{o}te d'Azur, CNRS, Laboratoire Lagrange, Bd de l'Observatoire, CS 34229, 06304 Nice cedex 4, France\label{aff96}
\and
Universit\'e Paris Cit\'e, CNRS, Astroparticule et Cosmologie, 75013 Paris, France\label{aff97}
\and
CNRS-UCB International Research Laboratory, Centre Pierre Binetruy, IRL2007, CPB-IN2P3, Berkeley, USA\label{aff98}
\and
Institut d'Astrophysique de Paris, 98bis Boulevard Arago, 75014, Paris, France\label{aff99}
\and
Aurora Technology for European Space Agency (ESA), Camino bajo del Castillo, s/n, Urbanizacion Villafranca del Castillo, Villanueva de la Ca\~nada, 28692 Madrid, Spain\label{aff100}
\and
Institut de F\'{i}sica d'Altes Energies (IFAE), The Barcelona Institute of Science and Technology, Campus UAB, 08193 Bellaterra (Barcelona), Spain\label{aff101}
\and
DARK, Niels Bohr Institute, University of Copenhagen, Jagtvej 155, 2200 Copenhagen, Denmark\label{aff102}
\and
Waterloo Centre for Astrophysics, University of Waterloo, Waterloo, Ontario N2L 3G1, Canada\label{aff103}
\and
Department of Physics and Astronomy, University of Waterloo, Waterloo, Ontario N2L 3G1, Canada\label{aff104}
\and
Perimeter Institute for Theoretical Physics, Waterloo, Ontario N2L 2Y5, Canada\label{aff105}
\and
Centre National d'Etudes Spatiales -- Centre spatial de Toulouse, 18 avenue Edouard Belin, 31401 Toulouse Cedex 9, France\label{aff106}
\and
Institute of Space Science, Str. Atomistilor, nr. 409 M\u{a}gurele, Ilfov, 077125, Romania\label{aff107}
\and
Consejo Superior de Investigaciones Cientificas, Calle Serrano 117, 28006 Madrid, Spain\label{aff108}
\and
Universidad de La Laguna, Departamento de Astrof\'{\i}sica, 38206 La Laguna, Tenerife, Spain\label{aff109}
\and
Dipartimento di Fisica e Astronomia "G. Galilei", Universit\`a di Padova, Via Marzolo 8, 35131 Padova, Italy\label{aff110}
\and
INFN-Padova, Via Marzolo 8, 35131 Padova, Italy\label{aff111}
\and
Institut f\"ur Theoretische Physik, University of Heidelberg, Philosophenweg 16, 69120 Heidelberg, Germany\label{aff112}
\and
Institut de Recherche en Astrophysique et Plan\'etologie (IRAP), Universit\'e de Toulouse, CNRS, UPS, CNES, 14 Av. Edouard Belin, 31400 Toulouse, France\label{aff113}
\and
Universit\'e St Joseph; Faculty of Sciences, Beirut, Lebanon\label{aff114}
\and
Departamento de F\'isica, FCFM, Universidad de Chile, Blanco Encalada 2008, Santiago, Chile\label{aff115}
\and
Universit\"at Innsbruck, Institut f\"ur Astro- und Teilchenphysik, Technikerstr. 25/8, 6020 Innsbruck, Austria\label{aff116}
\and
Institut d'Estudis Espacials de Catalunya (IEEC),  Edifici RDIT, Campus UPC, 08860 Castelldefels, Barcelona, Spain\label{aff117}
\and
Satlantis, University Science Park, Sede Bld 48940, Leioa-Bilbao, Spain\label{aff118}
\and
Institute of Space Sciences (ICE, CSIC), Campus UAB, Carrer de Can Magrans, s/n, 08193 Barcelona, Spain\label{aff119}
\and
Instituto de Astrof\'isica e Ci\^encias do Espa\c{c}o, Faculdade de Ci\^encias, Universidade de Lisboa, Tapada da Ajuda, 1349-018 Lisboa, Portugal\label{aff120}
\and
Cosmic Dawn Center (DAWN)\label{aff121}
\and
Niels Bohr Institute, University of Copenhagen, Jagtvej 128, 2200 Copenhagen, Denmark\label{aff122}
\and
Universidad Polit\'ecnica de Cartagena, Departamento de Electr\'onica y Tecnolog\'ia de Computadoras,  Plaza del Hospital 1, 30202 Cartagena, Spain\label{aff123}
\and
Kapteyn Astronomical Institute, University of Groningen, PO Box 800, 9700 AV Groningen, The Netherlands\label{aff124}
\and
Infrared Processing and Analysis Center, California Institute of Technology, Pasadena, CA 91125, USA\label{aff125}
\and
Dipartimento di Fisica e Scienze della Terra, Universit\`a degli Studi di Ferrara, Via Giuseppe Saragat 1, 44122 Ferrara, Italy\label{aff126}
\and
Istituto Nazionale di Fisica Nucleare, Sezione di Ferrara, Via Giuseppe Saragat 1, 44122 Ferrara, Italy\label{aff127}
\and
INAF - Osservatorio Astronomico di Brera, via Emilio Bianchi 46, 23807 Merate, Italy\label{aff128}
\and
INAF-Osservatorio Astronomico di Brera, Via Brera 28, 20122 Milano, Italy, and INFN-Sezione di Genova, Via Dodecaneso 33, 16146, Genova, Italy\label{aff129}
\and
ICL, Junia, Universit\'e Catholique de Lille, LITL, 59000 Lille, France\label{aff130}
\and
ICSC - Centro Nazionale di Ricerca in High Performance Computing, Big Data e Quantum Computing, Via Magnanelli 2, Bologna, Italy\label{aff131}
\and
Instituto de F\'isica Te\'orica UAM-CSIC, Campus de Cantoblanco, 28049 Madrid, Spain\label{aff132}
\and
CERCA/ISO, Department of Physics, Case Western Reserve University, 10900 Euclid Avenue, Cleveland, OH 44106, USA\label{aff133}
\and
Laboratoire Univers et Th\'eorie, Observatoire de Paris, Universit\'e PSL, Universit\'e Paris Cit\'e, CNRS, 92190 Meudon, France\label{aff134}
\and
Departamento de F{\'\i}sica Fundamental. Universidad de Salamanca. Plaza de la Merced s/n. 37008 Salamanca, Spain\label{aff135}
\and
Universit\'e de Strasbourg, CNRS, Observatoire astronomique de Strasbourg, UMR 7550, 67000 Strasbourg, France\label{aff136}
\and
Center for Data-Driven Discovery, Kavli IPMU (WPI), UTIAS, The University of Tokyo, Kashiwa, Chiba 277-8583, Japan\label{aff137}
\and
California Institute of Technology, 1200 E California Blvd, Pasadena, CA 91125, USA\label{aff138}
\and
Department of Physics \& Astronomy, University of California Irvine, Irvine CA 92697, USA\label{aff139}
\and
Department of Mathematics and Physics E. De Giorgi, University of Salento, Via per Arnesano, CP-I93, 73100, Lecce, Italy\label{aff140}
\and
INFN, Sezione di Lecce, Via per Arnesano, CP-193, 73100, Lecce, Italy\label{aff141}
\and
INAF-Sezione di Lecce, c/o Dipartimento Matematica e Fisica, Via per Arnesano, 73100, Lecce, Italy\label{aff142}
\and
Departamento F\'isica Aplicada, Universidad Polit\'ecnica de Cartagena, Campus Muralla del Mar, 30202 Cartagena, Murcia, Spain\label{aff143}
\and
Instituto de F\'isica de Cantabria, Edificio Juan Jord\'a, Avenida de los Castros, 39005 Santander, Spain\label{aff144}
\and
CEA Saclay, DFR/IRFU, Service d'Astrophysique, Bat. 709, 91191 Gif-sur-Yvette, France\label{aff145}
\and
Department of Computer Science, Aalto University, PO Box 15400, Espoo, FI-00 076, Finland\label{aff146}
\and
Instituto de Astrof\'\i sica de Canarias, c/ Via Lactea s/n, La Laguna 38200, Spain. Departamento de Astrof\'\i sica de la Universidad de La Laguna, Avda. Francisco Sanchez, La Laguna, 38200, Spain\label{aff147}
\and
Ruhr University Bochum, Faculty of Physics and Astronomy, Astronomical Institute (AIRUB), German Centre for Cosmological Lensing (GCCL), 44780 Bochum, Germany\label{aff148}
\and
Department of Physics and Astronomy, Vesilinnantie 5, 20014 University of Turku, Finland\label{aff149}
\and
Serco for European Space Agency (ESA), Camino bajo del Castillo, s/n, Urbanizacion Villafranca del Castillo, Villanueva de la Ca\~nada, 28692 Madrid, Spain\label{aff150}
\and
ARC Centre of Excellence for Dark Matter Particle Physics, Melbourne, Australia\label{aff151}
\and
Centre for Astrophysics \& Supercomputing, Swinburne University of Technology,  Hawthorn, Victoria 3122, Australia\label{aff152}
\and
Department of Physics and Astronomy, University of the Western Cape, Bellville, Cape Town, 7535, South Africa\label{aff153}
\and
DAMTP, Centre for Mathematical Sciences, Wilberforce Road, Cambridge CB3 0WA, UK\label{aff154}
\and
Kavli Institute for Cosmology Cambridge, Madingley Road, Cambridge, CB3 0HA, UK\label{aff155}
\and
Department of Astrophysics, University of Zurich, Winterthurerstrasse 190, 8057 Zurich, Switzerland\label{aff156}
\and
Department of Physics, Centre for Extragalactic Astronomy, Durham University, South Road, Durham, DH1 3LE, UK\label{aff157}
\and
IRFU, CEA, Universit\'e Paris-Saclay 91191 Gif-sur-Yvette Cedex, France\label{aff158}
\and
Oskar Klein Centre for Cosmoparticle Physics, Department of Physics, Stockholm University, Stockholm, SE-106 91, Sweden\label{aff159}
\and
Astrophysics Group, Blackett Laboratory, Imperial College London, London SW7 2AZ, UK\label{aff160}
\and
Univ. Grenoble Alpes, CNRS, Grenoble INP, LPSC-IN2P3, 53, Avenue des Martyrs, 38000, Grenoble, France\label{aff161}
\and
INAF-Osservatorio Astrofisico di Arcetri, Largo E. Fermi 5, 50125, Firenze, Italy\label{aff162}
\and
Dipartimento di Fisica, Sapienza Universit\`a di Roma, Piazzale Aldo Moro 2, 00185 Roma, Italy\label{aff163}
\and
Centro de Astrof\'{\i}sica da Universidade do Porto, Rua das Estrelas, 4150-762 Porto, Portugal\label{aff164}
\and
HE Space for European Space Agency (ESA), Camino bajo del Castillo, s/n, Urbanizacion Villafranca del Castillo, Villanueva de la Ca\~nada, 28692 Madrid, Spain\label{aff165}
\and
Dipartimento di Fisica - Sezione di Astronomia, Universit\`a di Trieste, Via Tiepolo 11, 34131 Trieste, Italy\label{aff166}
\and
Department of Astrophysical Sciences, Peyton Hall, Princeton University, Princeton, NJ 08544, USA\label{aff167}
\and
Theoretical astrophysics, Department of Physics and Astronomy, Uppsala University, Box 515, 751 20 Uppsala, Sweden\label{aff168}
\and
Minnesota Institute for Astrophysics, University of Minnesota, 116 Church St SE, Minneapolis, MN 55455, USA\label{aff169}
\and
Mathematical Institute, University of Leiden, Einsteinweg 55, 2333 CA Leiden, The Netherlands\label{aff170}
\and
Institute of Astronomy, University of Cambridge, Madingley Road, Cambridge CB3 0HA, UK\label{aff171}
\and
Space physics and astronomy research unit, University of Oulu, Pentti Kaiteran katu 1, FI-90014 Oulu, Finland\label{aff172}
\and
Department of Physics and Astronomy, Lehman College of the CUNY, Bronx, NY 10468, USA\label{aff173}
\and
American Museum of Natural History, Department of Astrophysics, New York, NY 10024, USA\label{aff174}
\and
Center for Computational Astrophysics, Flatiron Institute, 162 5th Avenue, 10010, New York, NY, USA\label{aff175}
\and
Department of Physics and Astronomy, University of British Columbia, Vancouver, BC V6T 1Z1, Canada\label{aff176}}    

%
%
%
%


%

%
\abstract{
We present a search for strong gravitational lenses in \Euclid imaging with high stellar velocity dispersion ($\sigmav>180\,\kms$) reported by SDSS and DESI. We performed expert visual inspection and classification of $11\,660$ \Euclid images. We discovered 38 grade A and 40 grade B candidate lenses, consistent with an expected sample of $\sim$32. Palomar spectroscopy confirmed 5 lens systems, while DESI spectra confirmed one, provided ambiguous results for another, and help to discard one. The \Euclid automated lens modeler modelled 53 candidates, confirming 38 as lenses, failing to model 9, and ruling out 6 grade B candidates. For the remaining 25 candidates we could not gather additional information. More importantly, our expert-classified non-lenses provide an excellent training set for machine learning lens classifiers. We create high-fidelity simulations of \Euclid lenses by painting realistic lensed sources behind the expert tagged (non-lens) luminous red galaxies. This training set is the foundation stone for the \Euclid galaxy-galaxy strong lensing discovery engine.
}
%
%
%

   \titlerunning{Euclid Q1: Early strong lens candidates}
   \authorrunning{Euclid Collaboration: K.~Rojas et. al.}
   
   \maketitle
%
%
%

%

%

\section{\label{sc:Intro}Introduction}

Strong gravitational lenses are powerful tools for understanding the most fundamental questions in astrophysics. They can be used study key insights into galaxy structure and cosmology \citep{Shajib2020,Treu2022}, constrain the nature of gravity \citep{Collett_2018}, the expansion history of our Universe \citep{Wells_2024} and the most massive galaxies within it \citep{Auger2009, Sonnenfeld2024}. Unfortunately, strong lenses are also rare. The typical deflection angle produced by a massive galaxy, assuming a spherical isothermal profile, is $\ang{;;1}$ so that strong lensing is only observed if a background galaxy is less than this angular distance from the optical axis between the observer and the deflector.

The first multiply imaged gravitationally lensed quasar was discovered in 1979 \citep{Walsh1979} and since then $\sim$\,10\,000 cases of strong gravitational lensing candidates by galaxies have been detected, with examples of lensed galaxies \citep{Jacobs2019,Petrillo2019,Canameras2020,Li2020,Rojas+2022,Savary+22,AcevedoBarroso24}, supernovae \citep{Kelly2015,Goobar2017,Pierel_2024} and even individual stars \citep{Kelly2018,Welch2022,Meena_2023} now known. This heterogeneous sample has allowed for a wide range of science, but has limited the statistical power of strong lensing. 

The main barrier to expanding the sample of strong gravitational lenses is the need for high angular resolution over a wide area of sky. Most galaxy lenses in the Universe have an Einstein radius of $\sim$\,0\farcs5 \citep{Collett2015} and so ground-based surveys (with seeing $\sim\ang{;;1}$) can only hope to resolve multiple imaging around the most massive galaxies. Observing from space provides the angular resolution to resolve more typical galaxy-scale lenses with $\sim$\,10 lenses discoverable per square degree in \HST imaging \citep{Faure_2008, Garvin_2022}. The VIS instrument \citep{EuclidSkyVIS} on \Euclid \citep{EuclidSkyOverview} will provide space-based imaging of over $14\,000$ deg$^2$, and so offers a step change in strong lens discovery potential. Forecasts by \citet{Collett2015} showed that \Euclid has the sensitivity to discover 170\,000 strong lenses. 

\Euclid will detect 1.5 billion unlensed galaxies, and so finding 170\,000 strong lenses will be a needle-in-a-haystack problem. Visually inspecting every galaxy will be impossible with such a large dataset, even though it has yielded large samples of lenses in smaller surveys \citep{Jackson2008,SpaceWarpsII, AcevedoBarroso24}. Machine learning has become a powerful tool for pre-selecting strong lens candidates \citep{Jacobs17,Jacobs2019,Petrillo2019, Li2020,Rojas+2022}, but even with a $99.99\%$ accurate classifier false positives would dominate. Currently citizen scientists \citep{SpaceWarpsI} or an even more accurate classifier will be needed to reduce the strong lens sample to a tractable problem.

The \cite{Q1cite} provides 63\,deg$^2$ of data representative of the full Euclid Wide Survey. This sample should contain $\sim600$ lenses \citep[scaling from][]{Collett2015} and gives us the first chance to implement, test, and validate our lens finding pipeline on a large scale. This paper is part of a series of papers developing, describing, and demonstrating the \Euclid strong lens discovery pipeline on the Q1 dataset \citep{Q1-TP001}.

This paper focuses on expert visual inspection of spectroscopically selected high velocity dispersion, massive galaxies as observed by the Dark Energy Spectrsoscopic Instrument (DESI; \citealp{DESI_EDR}) and the Sloan Digital Sky Survey (SDSS; \citealp{SDSSV_2019}). The cross section for strong gravitational lensing scales as velocity dispersion to the fourth power, and so focusing on massive galaxies maximises the chance of discovering new strong lenses before we train machine learning classifiers. Velocity dispersion and redshift are the key parameters for understanding the deflection angles produced by massive galaxies \citep{TreuKoopmans2004,Auger2009}, so results from the spectroscopic sample will be easier to interpret.

Starting with visual inspection of massive galaxies has three main benefits that enabled the machine learning discoveries made in \citet{Q1-SP048}. Firstly, it will provide a training set of expert vetted non-lenses and an expert classified sample of non-lens massive galaxies that can be used to create a positive training set by painting lensed sources behind them. Secondly, it will provide a small sample of real \Euclid lenses that, in addition to their important scientific value, can be used to validate the performance of our machine learning classifiers for recovering lenses in \Euclid data. Finally it will allow us to understand if \Euclid is delivering on the strong lensing forecast in \cite{Collett2015}.

The use of our simulated lenses to train machine learning classifiers is described in \citet{Q1-SP053}. The citizen science inspection pipeline is described in \citet{Q1-SP048}, where our main Q1 strong lens sample is also reported. We report our double source plane lens candidate sample in \citet{Q1-SP054}. In \citet{Q1-SP059} we present a machine learning and visual inspection ensemble analysis.

This paper is organised as follows:  In Sect.~\ref{sc:dataprep} we present how we selected the data used in this work, what we expect to find, and the design of the visual inspection including how we create the simulated test set. The results of the different visual inspection stages are in Sect.~\ref{sc:results} as well the analysis of the performance on the simulated set. In  Sect.~\ref{sc:spectroscopic} we present results from spectroscopic follow-up and in Sect.~\ref{sc:model} the results from automatic modelling. Finally in Sect.~\ref{sc:discussion} we present updates in the simulation pipeline and their implementation to build a training sample for train machine learning models and analyze the selection function. 

\section{\label{sc:dataprep} Data preparation and setup}  

In this section we present the design of our project, including the data selection, a forecast of what we should recover which takes into account the initial selection, the stages of the visual inspection procedure, and a description of the procedure to create simulations to evaluate the performance of the visual inspectors during the project.

\subsection{\label{sc:datasel} Spectroscopically selecting massive galaxies}
\begin{figure}[htbp!]
\centering
\includegraphics[angle=0,width=1.0\hsize]{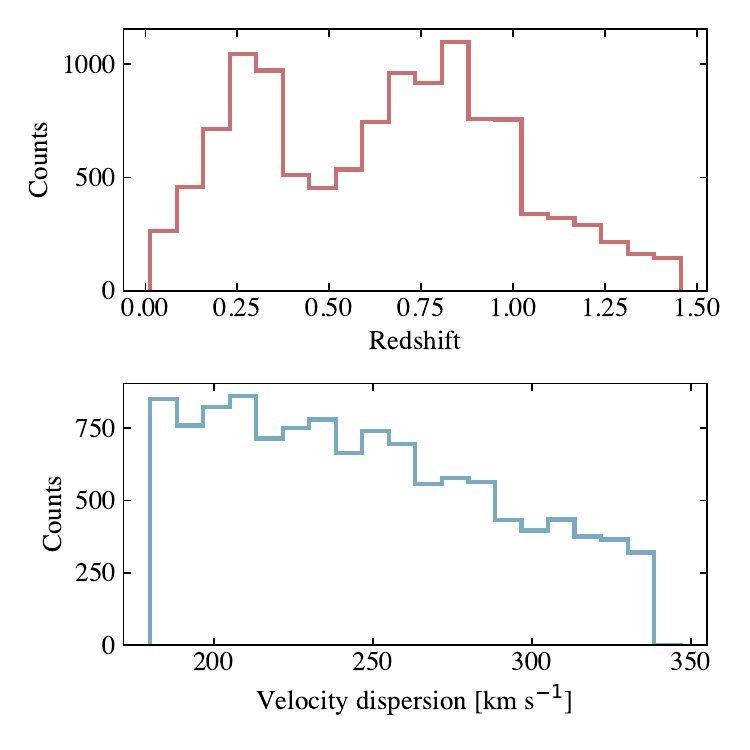}
\caption{Distribution of the redshift and velocity dispersion of the targets selected from DESI and SDSS with available \Euclid data for visual inspection. The distributions correspond to the pre-selection from two different survey and availability in \Euclid.}
\label{fig:zandvdisp}
\end{figure}

We selected massive galaxies with a velocity dispersion above $180\,{\rm km}\,{\rm s}^{-1}$  from DESI Early Data Release \citep[EDR,][]{DESI_EDR} and from SDSS Data Release 18 \citep[DR18,][]{SDSS_DR18}. In February 2024 we queried the \Euclid Science Archive System (SAS)  for any available product containing the selected targets. We found 11\,560 out of $\sim$\,290\,000 galaxies in the DESI sample and 100 out of 1.6 million in the SDSS sample. In Fig.~\ref{fig:zandvdisp}, we present the redshift and velocity dispersion distribution of the sample available at that query date. Most of the galaxies were found in the performance verification (PV) data. The majority are in the Euclid Deep Field North (EDF-N) and a few are part of the COSMOS-wide field. Hence, a few targets are outside the area covered by the Q1 release. For those targets, we present in this work the latest version available in the SAS and we call them pre-Q1 data. As we plan for a visual inspection, the difference in data processing between this and the final \Euclid Q1 data is not especially relevant. 

\subsection{\label{sc:forecast} Forecast of expected lenses}

By preselecting only the highest velocity dispersion galaxies in DESI and SDSS, we selected galaxies with large strong lensing cross sections. This means that the prevalence of lensing should be much higher than for randomly selected galaxies. \citet{Collett2015} used the {\sc{LensPop}} to forecast the expected number of lenses in the entire \Euclid survey to be 170\,000. This result is based on a population of singular isothermal ellipsoid (SIE) deflectors with velocity dispersions drawn from the observed velocity dispersion function of galaxies \citep{Choi_park_vogeley2007} uniformly distributed in comoving volume between $z=0$ and $z=2$. Behind these deflectors is a population of sources drawn from the LSST simulated catalogue \citep{Connoly2010}, with sources from redshift 0 to 5. 

We repurposed {\sc{LensPop}} to forecast the expected number of our spectroscopically selected objects that should be detectable as lenses with \Euclid. We replace the {\sc{LensPop}} deflector population with the observed redshift and velocity dispersions of our 11\,660 targets, assuming they are all SIEs. We retain the {\sc{LensPop}} background source population, simulation of \Euclid observations and the discoverable lens criteria used in \citet{Collett2015}, to which we refer the reader for further details. Applying this method we expect 32 lens systems should be discoverable from our 11\,660 targets. 

This estimate is far from perfect because it neglects any contribution to the lensing mass from groups and it assumes the DESI and SDSS velocity dispersions are correct, which is unlikely to be true for mergers. It also ignores any differences between our selection function and that in \citet{Collett2015}, who assumed a search on lens-subtracted \IE band images. We use \IE and infrared bands but do not subtract lens light. The statistical errors of $\sim$10\% on the velocity dispersion are irrelevant compared to these systematics.

This forecast also may not be accurate since it neglects the DESI and SDSS spectroscopic selection functions. The presence of a bright lensed arc will change the overall magnitude and colours of the system which may decrease (or increase) the probability of DESI or SDSS targeting the system. In summary, we should expect on the order of 30 lenses but it would not be surprising if the true number deviated by a factor of 5 in either direction.

\subsection{\label{sc:visualinpectiondesign} Visual inspection design}

To perform our visual inspection we used a slightly modified version of the visualisation tool developed by \cite{AcevedoBarroso24}. We used only the 1-by-1 sequential viewer, which displays a target cutout in the \IE band \citep{EuclidSkyVIS} and two colour composites using \IE-\YE-\HE and \YE-\JE-\HE \citep{EuclidSkyNISP}. We added or modified classification and subclassification buttons according to the different stages of this project. The final version of the classifier exhibits three main buttons for lensing classification: \textit{Lens}, \textit{Possible Lens} and \textit{Non Lens}, and six buttons for morphological classification: \textit{Merger}, \textit{Spiral}, \textit{Ring}, \textit{LRG}, \textit{Simulation} and \textit{Other}.

To achieve the goals of this project, we designed three stages. A beta stage for testing, and build a test set for following stage, stage-1, for detailed morphological classification, and stage-2 for lens grading. These stages are detailed as follows.

In the beta stage we aimed to test the modifications applied to the visualisation tool but also to build a small test set for the following stage. To do so, six visual inspectors classified 2000 random targets from the whole sample. Classifiers were asked to use all the buttons for testing purposes but the main focus was the identification of luminous red galaxies (LRGs) to build a test set containing simulated lens systems based on real images, as is explained in detail in Sect.~\ref{sc:sims} and LRGs as negative examples. From this stage we identified 700 LRGs without any lensing features. We then created simulations and kept a fraction of the LRGs as negative examples.  We constructed the test set such that visual inspectors should see a lens every 10--15 images. This does not represent the real rate of lenses ($\sim1$ in $1000$ galaxies) but could keep inspectors motivated.

In stage 1 we separated the sample into 6 groups with 5 visual inspectors in a group. Each person received a sample of about 2100 targets mixed with a test set of 200 labelled targets, 150 simulations and 50 LRGs, prepared with the information obtained in the beta test. This test set was the same for all individuals and had the purpose of detecting classifiers with poor completeness and purity that could bias the classification. The task in this stage was to do a detailed morphological classification, classifying each galaxy clicking in the respective button according to the categories: \textit{Merger}, \textit{Spiral}, \textit{Ring}, \textit{LRG}, and \textit{Other}, which are the most common contaminants in lens searches. In the option \textit{Other} we expected users to classify any other type of galaxy that is not listed in the options but also small galaxies where insufficient detail made accurate classification impossible. Furthermore, inspectors were instructed to identify lens candidates using the options: \textit{Lens} (L), \textit{Possible Lens} (PL) and \textit{Simulation}. We expected \textit{Lens} to be used for obvious lens systems and \textit{Possible Lens} for more doubtful ones, but no specific guidelines were given regarding the use of these buttons. With respect to the button \textit{Simulation}, this one was introduced for those who wanted to test their abilities to distinguish simulations from real lens systems, however its use was not mandatory and for final grading their votes counted as clicking \textit{Lens}.

Stage-2 was designed to grade all the possible lens systems. So here all visual inspectors re-inspected all targets that received at least one vote in the categories \textit{Lens} and \textit{Possible lens} during stage-1. Each inspector received the same set of data (the collection of \textit{Lens} and \textit{Possible lens}) along with a new test set, different from stage-1, and built with the labels collected in the first stage. This time the test set contains 111 simulations, so that the inspectors saw a lens every $\sim$\,10 images, and 80 non-lenses divided equally into four categories: LRGs, mergers, rings, and spirals. The purpose of this simulation set was not only to identify obtuse or random classifiers but also to evaluate the selection function. The simulations were carefully designed to almost evenly sample the parameter space of the Einstein radii and the S/N of the lensed images. In this stage the task was to classify the targets among: \textit{Lens} (L), \textit{Possible Lens} (PL), \textit{Non Lens} (NL) and \textit{Simulation}, this last category is optional. As in Stage-1 non-specific guidelines were given, but we expected that inspectors would click \textit{Lens} when an obvious lens system was displayed, \textit{Possible Lens} when the system may be a lens and \textit{Non Lens} when no sign of lensing features was present.

\subsection{\label{sc:cat_scoresystem} Catalogues and score system}

In order to create the final catalogues of galaxies in the categories \textit{Spirals}, \textit{Mergers} and \textit{Rings}, we kept any object that had a vote in the respective category from at least 3 out of the 4--5 inspectors. For LRGs we increased this cut to 2 votes out of 4--5, because usually LRGs do not get mistaken by any other category. Additional details into this morphological classification are in Sect.~\ref{sc:stage1}. 

In the case of lenses, we tried two score systems, a linear and a weighted one. In the linear one we assigned a linear score to the 3 categories, from 3 to 1 with: L=3, PL=2, NL=1, and then averaged among the number of participants. In the weighted one we counted the votes for \textit{Lens} 3 times more than the votes for \textit{Possible Lens}. For our particular case we observed that using the weighted score system shows a clearer separation in the sample. This results in a different scoring system than the one used in  \cite{Q1-SP048}. The equation to obtain the visual inspection score of each target is then:
\begin{equation}
\label{eqn:viscore}
\text{VI score} = \frac{3 \, N_{\rm L}  + 1 \, N_{\rm PL}}{\text{Total number of votes}}\,,
\end{equation}
where $N_{\rm L}$ is the number of votes in the \textit{Lens} category and $N_{\rm PL}$ in the \textit{Possible Lens} category. Using this scoring system, each lens received a unique score between 3 and 0. That is, the higher the score, the more confident inspectors were that the system is a lens. For the final lens catalogue we decided to make two cuts in the scores to separate the candidates into two groups. Category A, for a group of candidates mainly comprising obvious lens systems, with clear lens features, being the most voted by the inspectors. Category B, for a group of candidates with more doubtful lens systems. Any target that did not pass the two cuts was discarded. The VI score thresholds for these categories are discussed in Sect.~\ref{sc:stage2}.

\subsection{\label{sc:sims} Simulations}

To create the simulations we used all four \Euclid bands following the procedure in \cite{Rojas+2022} and using {\tt Lenstronomy}\footnote{\url{https://github.com/lenstronomy/lenstronomy}} \citep{Birrer2018,Birrer2021}. 

Our deflectors are selected LRGs with known-redshifts and velocity dispersions. We fit a S\'ersic profile to the \JE band image to obtain the ellipticity and central position of the galaxy, we will use these parameters to create our mass model. To minimise the log-likelihood in this fitting procedure we used a downhill simplex optimization \citep{nelder_mead_1965}  with 500 maximum iterations. We are not interested in a perfect fit, but in a rough and fast estimation, allowing some errors that could lead to a more diverse population of lenses. 

We selected sources to act as background galaxies from the HST/ACS F814W high-resolution \citep{Leauthaud2007,Scoville2007,Koekemoer2007} catalogue compiled by  \cite{Canameras2020}. These are HST/HSC combined sources, where the image is from HST and the colour information comes from Hyper Suprime Cam (HSC) ultra-deep stack images \citep{Aihara2018}. In this case we use the HST image and we assigned a similar magnitude to match the \Euclid filters. In the case of \IE band we used a combination of images with HSC $r$- band + $i$-band magnitudes. To match the infrarred bands we used the \cite{Ilbert_2009} catalogue to assign infrared magnitudes to our source galaxies. To do so we find the source with the nearest $gri$ magnitudes to ours and we assign their infrared magnitudes. In this case the closest infrared available filters in the catalogue to \Euclid \YE, \JE, and \HE filters are $z$-, $J$-, and $K$-bands respectively. This resulted in some cases with obvious mismatch colours when displayed in colour composite images, e.g. purple-ish lensing features. 

When both lens and source data are ready we match them in a way to ensure they will form Einstein radii greater than 0\farcs5. To do so we calculate the minimum redshift that a source should have to produce an Einstein radii of 0\farcs5 and we select a random source among all the ones with redshift above this value. We do not constrain the maximum Einstein radii, as we rarely form a system with such a large separation, allowing for this to happen. 

\begin{figure}[htbp!]
\centering
\includegraphics[angle=0,width=1.0\hsize]{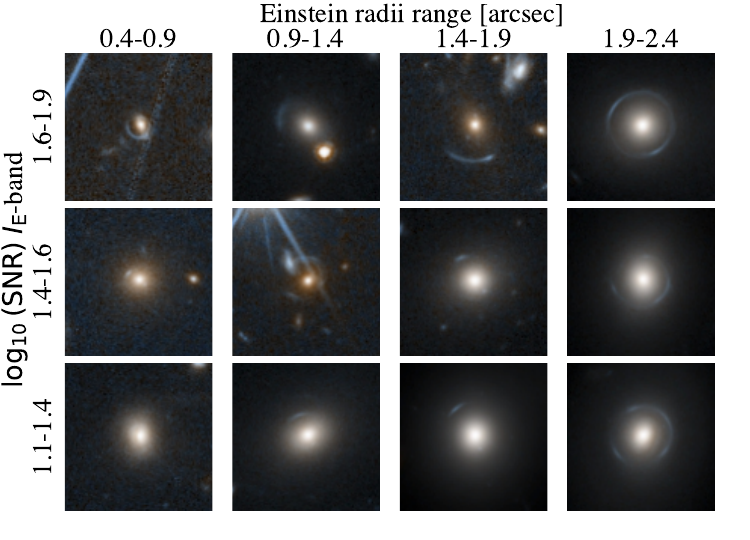}
\caption{Twelve example simulations selected to span different Einstein radii and log$_{10}$(SNR) in \IE band. Each cutout has a size of $10\arcsec \times 10\arcsec$ displayed using an MTF function using \IE and \YE bands.} 
\label{fig:sims}
\end{figure}

Once we have a lens-source pair we create an SIE mass model, whose parameters are the Einstein radius, $\theta_\sfont{E}$, position angle, the axis ratio and the central position. We derive the Einstein radius using the lens and source redshifts and the lens velocity dispersion. The position angle, the axis ratio and the central position are obtained from the S\'ersic profile fitted to the lens. We used this mass model to lens the background source light whose position is randomly selected within a square enclosing the caustics.  We downsampled the lensed source image to match lens pixel size. Then, we convolved the image with a Gaussian with a FWHM of 0\farcs15 for images in \IE filter or 0\farcs3 for those in the infrared, to broadly mimic the effect that the telescope PSF could produce, although these values do not match the exact FWHM of the PSFs in each filter. Finally, we re-scale the flux to the lens-image values. In Fig.~\ref{fig:sims} we show examples of simulations ranging over different combinations of Einstein radii and signal-to-noise ratio (SNR) of the source galaxy in the \IE image. The SNR is calculated taking the maximum value of the quotient between the cumulative sum of the pixels in the lensed source image before adding it to the lens galaxy image and the cumulative sum of the root mean square of the background standard deviation in the simulated image. The images are displayed using the midtone-transfer function \citep[MTF; see][]{Q1-SP048}.

\section{\label{sc:results} Results}

Our visual inspection had three stages. A beta-test and two main steps: stage-1 and stage-2. In this section we present the results found during these two main stages. 

\subsection{\label{sc:stage1} Stage 1: Morphological visual classification}
In stage-1 a total of 28 experts subscribed to perform the visual inspection, they were divided into 6 groups, 4 groups of 5 classifiers and 2 groups of 4 classifiers. We made sure each group had at least one experienced classifier, a person who had participated in several visual inspections before, to prevent doubtful or pessimistic classifiers from biasing the sample. The inspectors had three weeks to complete the task; after the deadline, 25 experts returned classifications. To ensure at least 4--5 classifications per group, classifier K.R.\ inspected 3 additional batches of data, keeping the original split of 4 groups with 5 classifications and 2 groups with 4 classifications. 

Based on the test set, the performance of all classifiers varied in completeness above 50$\%$ and purity above 97$\%$, except for one, whose completeness and purity were both below 50$\%$. Therefore, we decided not to use the classifications of this user, leaving us in the end with 3 groups with 5 classification and 3 groups with 4 classifications.

\begin{figure*}[htbp!]
  \begin{center}
    \includegraphics[angle=0,width=2.\columnwidth,]{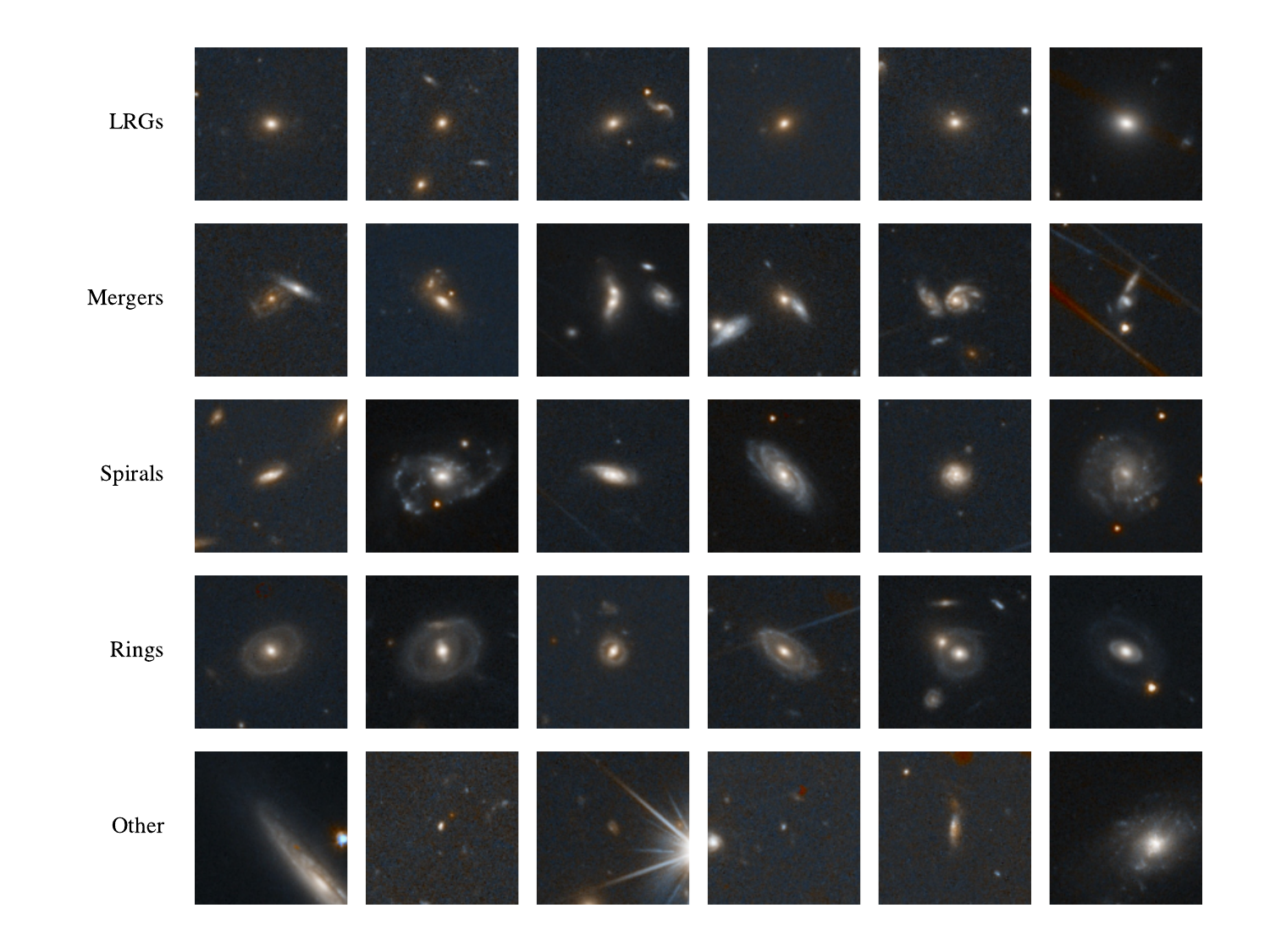}
   \end{center}
\caption{Six examples of targets classified in the LRG, mergers, spirals, rings, and other categories during stage 1 of the visual inspection. Cutouts have a size of $15\arcsec \times 15\arcsec$ and they are displayed using an MTF function using \IE and \YE bands.}
\label{fig:morph}
\end{figure*}

To analyse the morphological classification we counted how many votes in each category a target received. To consider a target to be in the categories: \textit{LRG}, \textit{Spiral}, \textit{Merger}, \textit{Ring} and \textit{Other}, we applied the following requirements: the targets must have at least three votes in the corresponding category and the target should have no vote in a lensing related category. In the case of LRGs, as we want a very clean sample to use them for simulations, we added the additional restriction that it should not have any votes in one of the other categories, removing possible confusing targets. As a result, we obtained 2578 spirals, 250 merges, 61 rings and 2477 galaxies in the category \textit{Others}. In the case of LRGs, 16$\%$ of the sample that complied with the general requirements did not pass the additional restriction leaving a sample of 2798 secure LRGs. Only 0.7$\%$ of the whole sample did not receive any classification in any category by any user, the main cause of this were targets missing \IE band information or artifacts in the image that do not allow a proper classification. 23$\%$ of the sample received confusing results not reaching a minimum of three votes in one category. These targets were not considered further. Examples of the best classified targets in these categories are shown in Fig.\,\ref{fig:morph}. One remark regarding the category \textit{Other} and \textit{Spiral} was noted in a post-classification survey, where some inspectors mentioned that they classified edge-on spirals in spiral and other inspectors in the category Others, so this type of galaxy can be found mixed in these two categories. 

Regarding strong lensing candidates, 1076 targets received at least one vote in one of the lensing related categories, including 14 targets with at least 3 votes as \textit{Lens} and 84 as \textit{Possible lens}. Interestingly 34 real targets were flagged by at least one person in the option \textit{Simulation}.

\subsection{\label{sc:stage2} Stage 2: Lens candidate grading}

All visual inspectors who completed stage-1 were invited to participate in stage-2. After two weeks, all but one returned classifications of all targets. In this stage we reclassified, only using the lensing related options, the 1076 targets with at least one vote in a lensing category from stage-1.

The user performance was evaluated using a different test set than the one used in stage 1. This updated set included simulations made with the previously classified LRGs and examples of different false positives. Most visual inspectors achieved a purity above 95$\%$ and a completeness above 70$\%$. However, three classifiers presented a purity below 80$\%$ and one had a completeness below 50$\%$. Consequently, we decided to exclude the classifications of these four visual inspectors from our final analysis. 

\begin{figure*}[htbp!]
\centering
\includegraphics[angle=0,width=1.0\hsize]{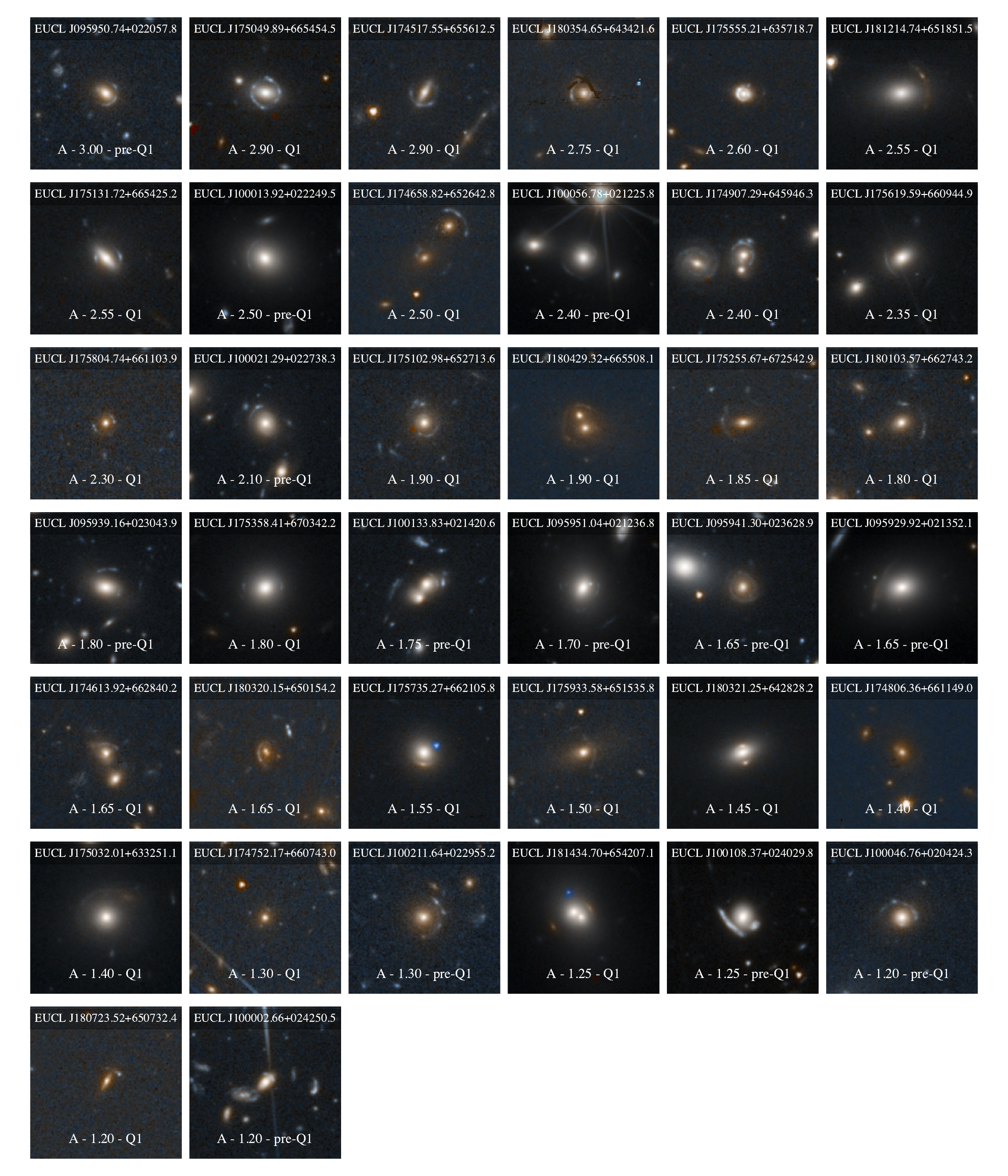}
\caption{Lens candidates in category A. Each image display on top the lens candidate name and on bottom the category, VI score, and data release. Each cutout has a size of $15\arcsec \times 15\arcsec$ and they are displayed using an MTF function using \IE and \YE bands.}
\label{fig:catA}
\end{figure*}

\begin{figure*}[htbp!]
\centering
\includegraphics[angle=0,width=1.0\hsize]{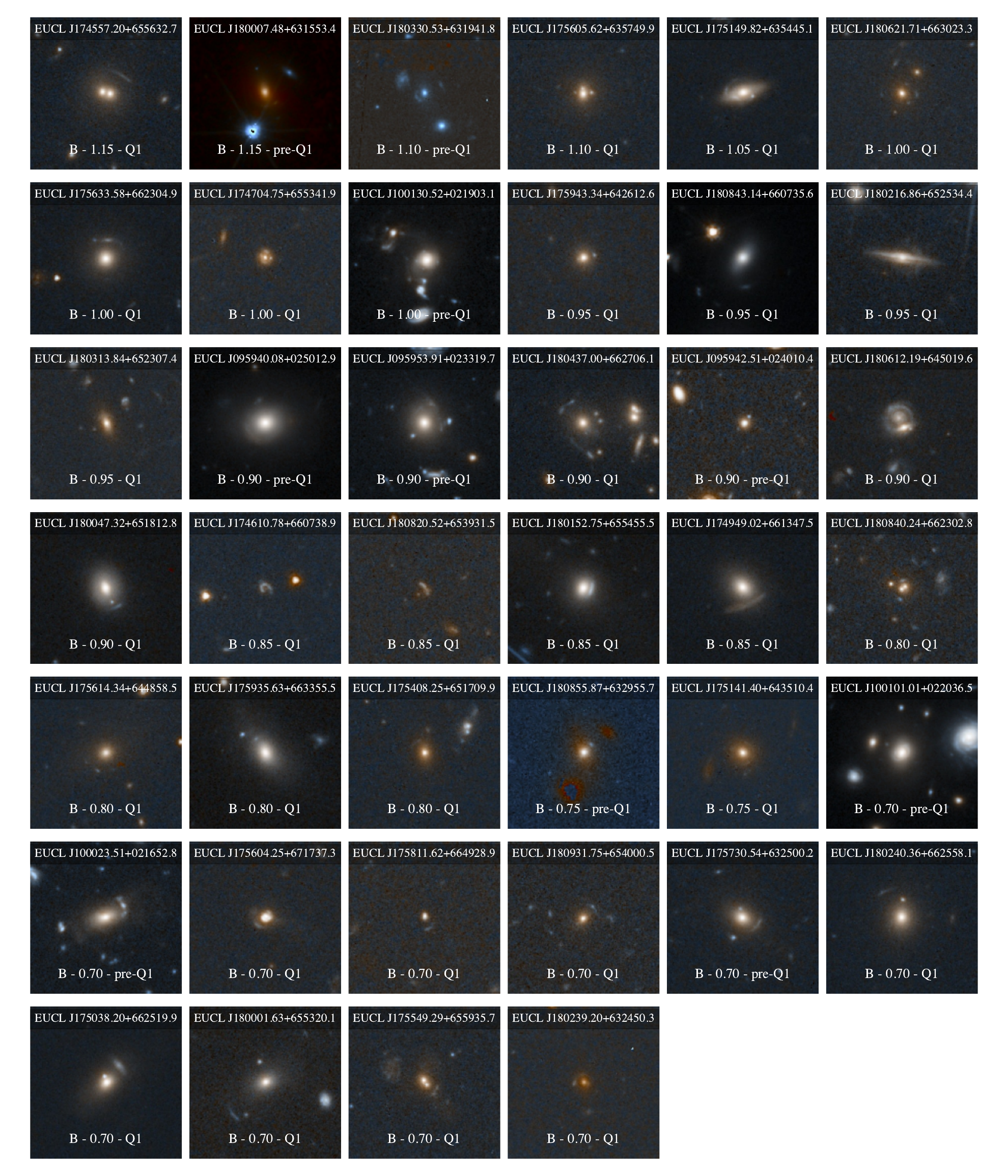}
\caption{Lens candidates in category B. Characteristics of the images are the same as Fig.\,\ref{fig:catA}.}
\label{fig:catB}
\end{figure*}

We calculated individual scores for each target following Eq.~(\ref{eqn:viscore}). By plotting all targets and their scores we visually decided to separate targets into 3 categories: A, B and Non-lens. The distinction between A and B can be seen as targets in category A are almost secure lens systems while in category B we have possible lens candidates and a few contaminants. For category A we obtained 36 targets with a score above 1.20 and for category B, 40 targets with scores between 1.20 and 0.70. The remaining targets were discarded. In Fig~\ref{fig:catA}~and~\ref{fig:catB} we show all lens candidates separated by category, their score and data release availability (Q1 or pre-Q1). In Tables~\ref{tab:catA}~and~\ref{tab:catB} we present the list of candidates in each category, their names, coordinates, redshifts, velocity dispersion, visual inspection score and references to the discovery publication in case they were previously detected. 

\begin{table*}[ht!]
\centering
\caption{Lens candidates in Category A.}
\label{tab:catA}
\begin{tabular}{ccccccccc}
Name & RA & Dec & $z_{\text{Lens}}$ & $\sigma_{\text{v}}$ [${\rm km}\,{\rm s}^{-1}$] & VI score & Model\tablefootmark{a} & $\theta_{\text{E}} [\arcsec]$  & Discovery \\
\hline
\hline
EUCL\,J095950.74+022057.8 & 149.961433 & \phantom{6}2.349415 & 0.94 & 331 & 3.00 & NM  & -- &  [1]\\
EUCL\,J174517.55+655612.5 & 266.323138 & 65.936820 & 0.61 & 300 & 2.90 & Y/Y & 1.23 &  This work \\
EUCL\,J175049.89+665454.5 & 267.707904 & 66.915153 & 0.38 & 200 & 2.90 & Y/Y & 1.39 &  This work \\
EUCL\,J180354.65+643421.6 & 270.977718 & 64.572670 & 0.52 & 193 & 2.75 & Y/Y & 1.06 &  This work \\
EUCL\,J175555.21+635718.7 & 268.980054 & 63.955196 & 0.33 & 219 & 2.60 & Y/Y & 0.53 &  This work \\
EUCL\,J181214.74+651851.5 & 273.061422 & 65.314333 & 0.31 & 248 & 2.55 &  Y/Y & 1.73  &  This work \\
EUCL\,J175131.72+665425.2 & 267.882204 & 66.907018 & 0.26 & 333 & 2.55 & Y/Y & 1.20 &  This work \\
EUCL\,J100013.92+022249.5 & 150.058040 & \phantom{6}2.380438 & 0.35 & 227 & 2.50 &   NM & -- &  [2] \\
EUCL\,J174658.82+652642.8 & 266.745111 & 65.445234 & 0.81 & 204 & 2.50 & N/- & -- &  This work \\
EUCL\,J174907.29+645946.3 & 267.280382 & 64.996215 & 0.48 & 298 & 2.40 & Y/Y & 0.91  &  This work \\
EUCL\,J100056.78+021225.8 & 150.236610 & \phantom{6}2.207190 & 0.36 & 249 & 2.40 &  NM  & -- &  [2] \\
EUCL\,J175619.59+660944.9 & 269.081656 & 66.162488 & 0.27 & 245 & 2.35 & Y/Y & 1.28 &  This work \\
EUCL\,J175804.74+661103.9 & 269.519782 & 66.184429 & 0.92 & 232 & 2.30 & Y/Y & 0.99 &  This work \\
EUCL\,J100021.29+022738.3 & 150.088725 & \phantom{6}2.460639 & 0.73 & 221 & 2.10 &  NM & -- & This work \\
EUCL\,J175102.98+652713.6 & 267.762422 & 65.453784 & 0.73 & 312 & 1.90 & Y/Y & 1.45 &  This work \\
EUCL\,J180429.32+665508.1 & 271.122184 & 66.918930 & 0.67 & 335 & 1.90 & N/- & -- &  This work \\
EUCL\,J175255.67+672542.9 & 268.231987 & 67.428601 & 0.75 & 185 & 1.85 & Y/Y & 1.47 &  This work \\
EUCL\,J095939.16+023043.9 & 149.913197 & \phantom{6}2.512212 & 0.72 & 234 & 1.80&   NM & -- & [3] \\
EUCL\,J175358.41+670342.2 & 268.493385 & 67.061738 & 0.20 & 256  & 1.80 & Y/Y & 1.70  &  This work \\
EUCL\,J180103.57+662743.2 & 270.264883 & 66.462022 & 0.67 & 243 & 1.80 & Y/Y & 1.80 &  This work \\
EUCL\,J100133.83+021420.6 & 150.390992 & \phantom{6}2.239058 & 0.67 & 221  & 1.75 & NM & -- &  This work \\
EUCL\,J095951.04+021236.8 & 149.962680 & \phantom{6}2.210235 & 0.42 & 228  & 1.70 & NM & -- &  This work \\
EUCL\,J095941.30+023628.9 & 149.922088 & \phantom{6}2.608045 & 0.89 & 263  & 1.65 & NM & -- &  This work \\
EUCL\,J095929.92+021352.1 & 149.874700 & \phantom{6}2.231164 & 0.34 & 216  & 1.65 & NM & -- &  This work \\
EUCL\,J174613.92+662840.2 & 266.558025 & 66.477847 & 0.63 & 248 & 1.65 & N/- & -- &  This work \\
EUCL\,J180320.15+650154.2 & 270.833961 & 65.031730 & 0.81 & 299  & 1.65 & Y/Y & 0.47 &  This work \\
EUCL\,J175735.27+662105.8 & 269.396979 & 66.351620 & 0.28 & 263 & 1.55 & Y/Y & 0.64 &  This work \\
EUCL\,J175933.58+651535.8 & 269.889925 & 65.259966 & 0.56 & 217 & 1.50 & Y/Y & 1.18 &  This work \\
EUCL\,J180321.25+642828.2 & 270.838560 & 64.474516 & 0.19 & 228 & 1.45 & Y/Y & 0.77 &  This work \\
EUCL\,J174806.36+661149.0 & 267.026514 & 66.196947 & 0.62 & 333 & 1.40 & Y/Y & 0.31  &  This work \\
EUCL\,J175032.01+633251.1 & 267.633377 & 63.547552 & 0.28 & 261 & 1.40 & Y/Y & 1.49 &  This work \\
EUCL\,J174752.17+660743.0 & 266.967395 & 66.128635 & 0.70 & 276  & 1.30 & Y/Y & 1.30  &  This work \\
EUCL\,J100211.64+022955.2 & 150.548511 & \phantom{6}2.498683 & 0.88 & 257  & 1.30 & NM & --  &  [3] \\
EUCL\,J181434.70+654207.1 & 273.644590 & 65.701988 & 0.20 & 277  & 1.25 & N/- & -- &  This work \\
EUCL\,J100108.37+024029.8 & 150.284904 & \phantom{6}2.674945 & 0.25 & 291  & 1.25 & NM & -- &  [1] \\
EUCL\,J100046.76+020424.3 & 150.194841 & \phantom{6}2.073427 & 0.94 & 234  & 1.20 & NM & -- &  This work \\
EUCL\,J180723.52+650732.4 & 271.848021 & 65.125671 & 0.89 & 310 & 1.20 & Y/Y & 1.08 &  This work \\
EUCL\,J100002.66+024250.5 & 150.011105 & \phantom{6}2.714039 & 0.74 & 259  & 1.20 & NM & -- &  This work \\
\hline
\end{tabular}
\tablefoot{
\tablefoottext{a}{In this column we report both model success/model expert evaluation. Where Y/N stands for yes and no regarding if the system was successfully modelled and if experts think the system is a lens based on the model. Systems with pre-Q1 data were not modelled, hence NM stands for No Model.}}
\tablebib{[1]~\cite{Pourrahmani2018}, [2]~\cite{Garvin_2022}, [3]~\cite{More2012}}
\end{table*}

\begin{table*}[ht!]
\centering
\caption{Lens candidates in category B.}
\label{tab:catB}
\begin{tabular}{ccccccccc}
Name & RA & Dec & $z_{\text{Lens}}$ & $\sigma_{\text{v}}$ [${\rm km}\,{\rm s}^{-1}$] & VI score & Model\tablefootmark{a} & $\theta_{\text{E}} [\arcsec]$  & Discovery \\
\hline
\hline
EUCL\,J174557.20+655632.7 & 266.488341 & 65.942430 & 0.60 & 218 & 1.15 &  N/- & -- & This work \\
EUCL\,J180007.48+631553.4 & 270.031189 & 63.264859 & 0.29 & 316 & 1.15 & NM  & -- & This work \\
EUCL\,J180330.53+631941.8 & 270.877209 & 63.328302 & 0.92 & 236 & 1.10 & NM  & -- & This work \\
EUCL\,J175605.62+635749.9 & 269.023418 & 63.963877 & 0.70 & 283 & 1.10 &  N/- & -- & This work \\
EUCL\,J175149.82+635445.1 & 267.957618 & 63.912530 & 0.33 & 191 & 1.05 & N/- & -- & This work \\
EUCL\,J180621.71+663023.3 & 271.590471 & 66.506485 & 0.85 & 280 & 1.00 & Y/Y & 1.43 & This work \\
EUCL\,J175633.58+662304.9 & 269.139948 & 66.384706 & 0.39 & 230 & 1.00 & Y/Y & 0.99 & This work \\
EUCL\,J174704.75+655341.9 & 266.769824 & 65.894996 & 0.77 & 236 & 1.00 & Y/Y & 0.54 & This work \\
EUCL\,J100130.52+021903.1 & 150.377170 & \phantom{6}2.317550 & 0.70 & 271 & 1.00 & NM  & -- & [1] \\
EUCL\,J175943.34+642612.6 & 269.930605 & 64.436846 & 0.65 & 253 & 0.95 & Y/Y & 0.75 & This work \\
EUCL\,J180843.14+660735.6 & 272.179785 & 66.126567 & 0.22 & 292 & 0.95 & Y/N & -- & This work \\
EUCL\,J180216.86+652534.4 & 270.570265 & 65.426239 & 0.48 & 283 & 0.95 & N/- & -- & This work \\
EUCL\,J180313.84+652307.4 & 270.807705 & 65.385405 & 0.77 & 311 & 0.95 & Y/Y & 0.73 & This work \\
EUCL\,J095940.08+025012.9 & 149.917010 & \phantom{6}2.836917 & 0.36 & 235 & 0.90 & NM & -- & This work \\
EUCL\,J095953.91+023319.7 & 149.974664 & \phantom{6}2.555488 & 0.73 & 254 & 0.90 & NM & -- & This work \\
EUCL\,J180437.00+662706.1 & 271.154194 & 66.451702 & 0.67 & 305 & 0.90 & Y/N & -- & This work \\
EUCL\,J095942.51+024010.4 & 149.927150 & \phantom{6}2.669560 & 0.79 & 324 & 0.90 & NM & -- & This work \\
EUCL\,J180612.19+645019.6 & 271.550799 & 64.838791 & 0.58 & 217 & 0.90 & N/- & -- & This work \\
EUCL\,J180047.32+651812.8 & 270.197198 & 65.303572 & 0.29 & 193 & 0.90 & Y/Y & 1.48 & This work \\
EUCL\,J174610.78+660738.9 & 266.544935 & 66.127475 & 1.00 & 267 & 0.85 & N/- & -- & This work \\
EUCL\,J180820.52+653931.5 & 272.085531 & 65.658777 & 0.52 & 330 & 0.85 & N/- & -- & This work \\
EUCL\,J180152.75+655455.5 & 270.469808 & 65.915421 & 0.36 & 278 & 0.85 & Y/N & -- & This work \\
EUCL\,J174949.02+661347.5 & 267.454275 & 66.229872 & 0.35 & 215 & 0.85 & Y/N & -- &This work \\
EUCL\,J180840.24+662302.8 & 272.167673 & 66.384137 & 0.92 & 245 & 0.80 & Y/Y & 0.92 & This work \\
EUCL\,J175614.34+644858.5 & 269.059778 & 64.816274 & 0.64 & 241 & 0.80 & Y/Y & 0.92 & This work \\
EUCL\,J175935.63+663355.5 & 269.898476 & 66.565430 & 0.39 & 243 & 0.80 & Y/Y & 0.87 & This work \\
EUCL\,J175408.25+651709.9 & 268.534402 & 65.286092 & 0.67 & 218 & 0.80 & Y/Y & 0.95 & This work \\
EUCL\,J180855.87+632955.7 & 272.232826 & 63.498831 & 0.64 & 275 & 0.75 & NM & -- & This work \\
EUCL\,J175141.40+643510.4 & 267.922538 & 64.586238 & 0.64 & 227 & 0.75 & Y/Y & 1.10 & This work \\
EUCL\,J100101.01+022036.5 & 150.254245 & \phantom{6}2.343489 & 0.60 & 311 & 0.70 & NM & -- & This work \\
EUCL\,J100023.51+021652.8 & 150.097985 & \phantom{6}2.281358 & 0.75 & 215 & 0.70 & NM & -- &  [2] \\
EUCL\,J175604.25+671737.3 & 269.017725 & 67.293715 & 0.69 & 311 & 0.70 & Y/Y & 0.86 & This work \\
EUCL\,J175811.62+664928.9 & 269.548440 & 66.824712 & 1.09 & 269 & 0.70 & Y/N & -- & This work \\
EUCL\,J180931.75+654000.5 & 272.382329 & 65.666818 & 0.93 & 188 & 0.70 & Y/Y & 1.14 & This work \\
EUCL\,J175730.54+632500.2 & 269.377274 & 63.416732 & 0.48 & 256 & 0.70 & NM & -- & This work \\
EUCL\,J180240.36+662558.1 & 270.668199 & 66.432832 & 0.48 & 270 & 0.70 & Y/N & -- & This work \\
EUCL\,J175038.20+662519.9 & 267.659169 & 66.422196 & 0.40 & 255 & 0.70 & Y/Y & 1.03 & This work \\
EUCL\,J180001.63+655320.1 & 270.006810 & 65.888924 & 0.68 & 281 & 0.70 & Y/Y & 0.94 & This work \\
EUCL\,J175549.29+655935.7 & 268.955380 & 65.993254 & 0.82 & 293 & 0.70 & Y/Y & 2.38 & This work \\
EUCL\,J180239.20+632450.3 & 270.663350 & 63.413974 & 1.05 & 186 & 0.70 & Y/Y & 0.72 & This work \\
\hline
\end{tabular}
\tablefoot{
\tablefoottext{a}{Same definitions as in Tab.~\ref{tab:catA}.}}
\tablebib{[1]~\cite{cao2020}, [2]~\cite{Pawase2014}}
\end{table*}

\section{\label{sc:spectroscopic} Spectroscopic follow-up}

In this section, we present the spectroscopic analysis of observations from the Palomar Observatory and the inspection on public available spectra from DESI and SDSS archives searching for emission or absorption lines at a redshift different from that of the reported lens, which could provide an estimate of the source redshift. 

\subsection{\label{sc:palomar_spectra} Palomar observations}

We obtained optical spectroscopic follow-up of 12 category A candidates in the EDFN using the Double Spectrograph 
\citep[DBSP,][]{Oke1982} on the 5m Hale telescope at Palomar Observatory between July and September 2024.  Table~\ref{tab:palomar_spectroscopy} presents the targets for which we were able to measure at least one redshift in the possible strong lens system.  The nights all had seeing ranging from 1\farcs1 to 1\farcs5; most observations were obtained with $\sim$\,1\farcs3 seeing.  Half the nights were photometric, meaning no cloud coverage, and the other half had variable levels of cloud coverage ranging from minimal to sufficiently severe and opaque monsoon clouds that the dome was shuttered.  For each source, we obtained two or three exposures of 1200\,s using the 1\farcs5 slit, the 600 line blue grating (blazed at 4000\,\AA), the 5500\,\AA\, dichroic, and the 316 line red grating (blazed at 7500\,\AA).  The slits were aligned on the candidate lensing galaxy at a position angle to cover the putative lensed source feature. The data were reduced using standard techniques within Image Reduction and Analysis Facility (IRAF), and the quality (Q) of spectroscopic redshifts were assessed as either quality A, implying multiple detected features and a highly secure redshift, or quality B, implying some ambiguity to the reported redshift either due to the robustness of the putative detected feature or ambiguity into the identification of that feature.

\begin{figure*}[htbp!]
\centering
\includegraphics[angle=0,width=1.0\hsize]{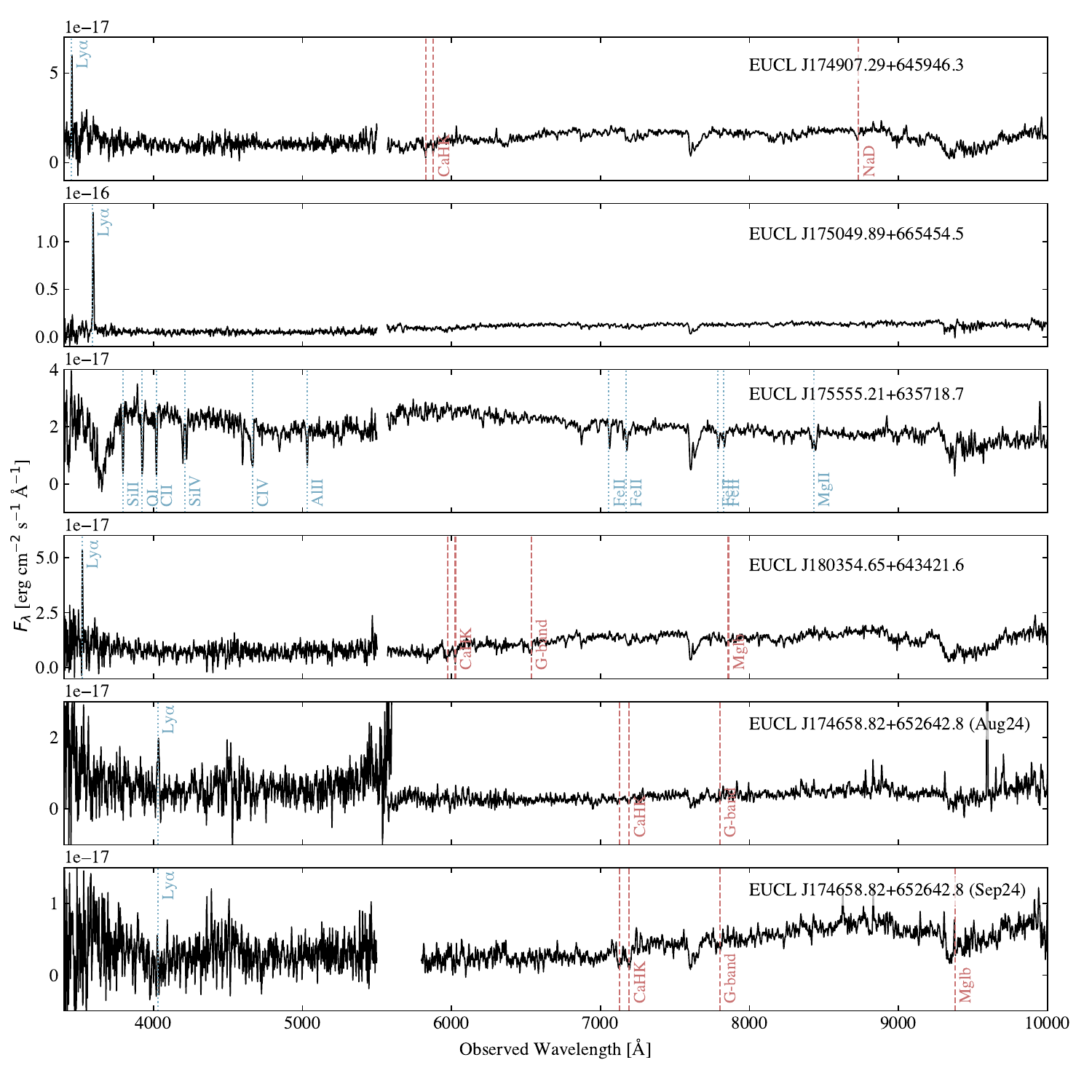}
\caption{Spectra of the five targets with source redshift estimations. Identified spectral lines are labeled, with emission lines indicated at the top of the image and absorption lines at the bottom. Lines associated with the lens galaxy are shown in red with a dashed style, while those corresponding to the source are in blue with a dotted style.}
\label{fig:palomar_spec}
\end{figure*}

All the lensing galaxies proved to be early-type galaxies with \ion{Ca}{II} H \& K absorption and, generally, strong 4000\,\AA\, breaks. We obtained quality A redshifts for four lensed sources, all at $z \sim 2$, as well as one quality B redshift at $z = 2.316$ (Fig.~\ref{fig:palomar_spec}).  In most cases, the lensed background source was revealed as a slightly offset or extended blue emission line coincident with the early-type lensing galaxy.  Since the emission features did not correspond to any strong, redshifted spectral features in early-type galaxies (which generally do not have emission lines), the most plausible identifications were lensed Ly$\alpha$ at $z \sim 2$. One lensed source, EUCL~J175555.21+635718.7, does not show Ly$\alpha$ emission but instead shows the classic spectrum of a Lyman-break galaxy with multiple absorption lines due to the interstellar medium. A detailed analysis and further follow-up of this target and EUCL\,J174907.29+645946.3, a possible double source plane candidate, will be presented in Moustakas et al. (in prep). 

\subsection{\label{sc:add_spectra} Additional available spectra}

We visually inspected the available DESI and SDSS spectra for all 78 targets. We found that all ten redshift for the lens galaxies obtained from Palomar are in agreement with the redshifts previously reported. Regarding the source detections made by Palomar, in most of the cases, the Ly$\alpha$ emission line is out of the DESI spectra coverage or very near to the edge making its detection in DESI data impossible or unreliable. Additional spectral features were identified in only four targets, including the EUCL\,J175555.21+635718.7, the Lyman-break galaxy previously mentioned. Based on insights from Palomar spectroscopic data, we believe that in many cases, the emission or absorption lines may fall outside the observed spectral range or near the edges, where noise levels are high, making detection challenging. Additionally, the integration time may not have been sufficient to capture the often faint signals from the sources. The findings for the three additional detected targets are described below.

For EUCL\,J174613.92+662840.2 we found an emission line at $8587\,\AA$. Based on its shape, it is likely  \ion{O}{II}, corresponding to a source redshift of 1.303.

In the spectra of EUCL\,J100101.01+022036.5 we identified weak emission features at $7217\,\AA$, $9603\,\AA$ and $9693\,\AA$, which could correspond to \ion{O}{II} and \ion{O}{III]} doublet indicating a source galaxy at $z\sim0.935$. However, due to the weakness of the signal, this detection remains ambiguous. 

The candidate EUCL\,J180152.75+655455.5 exhibits a clear set of emission lines at a different redshift than the absorptions lines corresponding to the lens ($z=0.36$). We identified \ion{O}{II}, H$\beta$, \ion{O}{III]} doublet and H$\alpha$ corresponding to a $z\sim0.48$. The close proximity of these two galaxies suggests that this system is not a strong lens candidate.

\begin{table*}[ht!]
\centering
\caption{Palomar spectroscopy of strong lens candidates.}
\begin{tabular}{lccccccc}
\hline
\hline
Name &  ObsDate (UT) & PA [deg] & $z_\mathrm{lens}$ & $z_\mathrm{source}$ & Q/Q\tablefootmark{a} & Notes \\
\hline
EUCL\,J174907.29+645946.3 &  2024 Aug 02 & $-15$ & 0.481 & 1.839 & A/A & Compound lens \\
EUCL\,J175049.89+665454.5 &  2024 Jul 10 & +150 &   & 1.956 & -/A &  \\
EUCL\,J175555.21+635718.7 &  2024 Jul 10 & +90 &   & 2.011 & -/A &  Lyman-break \\
EUCL\,J180354.65+643421.6 &  2024 Aug 02 & $-60$ & 0.518 & 1.897 & A/A &  \\
\hline
EUCL\,J174658.82+652642.8 &  2024 Aug 03 & 0 & 0.812 & 2.316 & B/B &   \\
                         &  2024 Sep 11 & $-15$ & 0.812 &   & A/- &   \\
\hline
EUCL\,J174517.55+655612.5 &  2024 Aug 02 & $-30$ & 0.611 &   & A/- &   \\
EUCL\,J175102.98+652713.6 &  2024 Aug 11 & +35 & 0.734 & & B/- &   \\
EUCL\,J175131.72+665425.2 &  2024 Aug 03 & +90 & 0.264 &  & A/- &  \\
EUCL\,J175619.59+660944.9 &  2024 Aug 11 & +120 & 0.271 &  & A/- &   \\
EUCL\,J175735.27+662105.8 &  2024 Aug 11 & 0 & 0.285 &  & A/- &   \\
EUCL\,J180321.25+642828.2 &  2024 Jul 10 & +10 & 0.186 &   & A/- &   \\
\hline
EUCL\,J181214.74+651851.5 &  2024 Aug 02 & $-40$ & 0.308 &   & A/- &   \\
                         &  2024 Sep 08 & $-40$ & 0.309 &   & A/- &   \\
\hline
\end{tabular}
\tablefoot{
\tablefoottext{a}{Quality of the redshift calculation for lens/source; see Sect.~\ref{sc:spectroscopic} for details.  }
}
\label{tab:palomar_spectroscopy}
\end{table*}

\section{\label{sc:model} Lens modeling}

The \Euclid strong lens modeling pipeline (Nightingale in prep.) was applied to the 53 lens candidates with Q1 data, 24 category A and 29 category B. This final step aims to provide insights to assess whether the candidates are potential strong lensing systems.

\subsection{Approach}

We perform automated strong lens modeling of all the candidates with Q1 available data using the \Euclid strong lens modelling pipeline\footnote{\href{https://github.com/Jammy2211/euclid\_strong\_lens\_modeling\_pipeline}{github.com/Jammy2211/euclid\_strong\_lens\_modeling\_pipeline}}, adaptated from the lens modelling software \texttt{PyAutoLens}\footnote{\href{https://github.com/Jammy2211/PyAutoLens}{github.com/Jammy2211/PyAutoLens}} \citep{Nightingale2021}. 

The lens mass is modelled as an isothermal profile
\begin{equation}
\label{eqn:SPLEkap}
\kappa (\xi) = \frac{1}{1 + q^{\rm mass}} \bigg( \frac{\theta^{\rm mass}_{\rm E}}{\xi} \bigg)\,,
\end{equation}
where $\theta^{\rm mass}_{\rm E}$ is the Einstein radius. Deflection angles are calculated using \citet{Tessore2015}'s method in {\tt PyAutoLens}. External shear is included, parameterized as $(\gamma_1^{\rm ext}, \gamma_2^{\rm ext})$, with the shear magnitude and orientation given by
\begin{equation}
    \label{eq:shear}
    \gamma^{\rm ext} = \sqrt{\gamma_{\rm 1}^{\rm ext^{2}}+\gamma_{\rm 2}^{\rm ext^{2}}}, \, \,
    \tan{2\phi^{\rm ext}} = \frac{\gamma_{\rm 2}^{\rm ext}}{\gamma_{\rm 1}^{\rm ext}}\,.
\end{equation}
The deflection angles due to the external shear are computed analytically.

 The \Euclid strong lens modelling pipeline models the lens galaxy's light using a multi-Gaussian expansion (MGE, \citealt{He2024}), accounts for PSF blurring, and subtracts this model from the observed image. A mass model (isothermal distribution) ray-traces image pixels to the source-plane, where a pixelized source reconstruction is performed using an adaptive Delaunay mesh. The pipeline iteratively fits various combinations of light, mass, and source models; the pipeline initially fits a simpler model using an MGE source for efficient and robust convergence towards accurate results, then subsequent stages employ the more complex Voronoi source reconstruction. The pipeline chains together five lens model fits in total. 

For further description of \texttt{PyAutoLens}, see \citet{He2024}, \citet{Nightingale2024}, and Nightingale (in prep.) for full details. We also provide more details in \cite{Q1-SP048} Appendix A.

\subsection{Modelling results}

The first step assessed whether the automated modelling was successful, based primarily on how well the model reproduced the observed lensed source emission. The critical curves of the mass model and the source plane were also evaluated. A successful lens model does not necessarily confirm the candidate as a strong lens but indicates that the model fit the data as expected. For instance, if the observed emission in the image-plane is singly imaged without a counter-image and the model reflects this, the fit is deemed successful, even though the candidate is not a strong lens. Overall, 44 out of 53 candidates ($83\%$) were successfully modelled.

Among the 44 successful fits, experts evaluated whether the candidates were genuine strong lenses based on the models. Of these, 38 were classified as strong lenses, while 6 were determined not to be. All six non-lens belonged to  category B, including EUCL\,J180152.75+655455.5, which was blindly ruled out by this pipeline. This decision was later supported by redshift estimation of the two galaxies (Sect.~\ref{sc:add_spectra}, $z1=0.36$ and $z2=0.48$), confirming that this is not a strong lensing interaction. In Tables\,\ref{tab:catA}~and~\,\ref{tab:catB} we present for each candidate the model success and the decision of the experts. When the model fits the system successfully and the experts agree that the candidate is a lens, we also report the Einstein radii ($\theta_{\text{E}}$).



\section{\label{sc:discussion} Discussion}

We have morphologically categorized about $5000$ galaxies, discovered around 70 lens candidates, conducted a spectroscopic campaign at the Palomar Observatory to confirm five of them, and successfully automatically modelled 44. In this section, we discuss the lensing selection function and how we used our results to build the training set used in  \citet{Q1-SP048} and   \citet{Q1-SP053}.

\subsection{\label{sc:selfunc} Lensing selection function}

\begin{figure}[htbp!]
\centering
\includegraphics[angle=0,width=1.0\hsize]{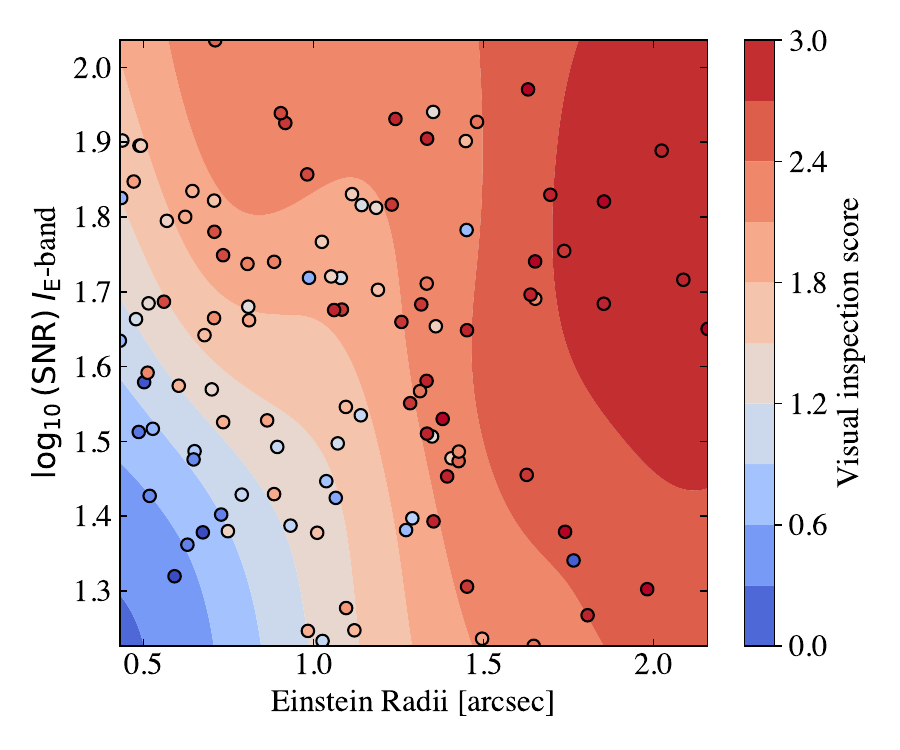}
\caption{Visual inspection score related to the Einstein radii and log$_{10}$(SNR) of the lensed source in \IE band. Colour maps represent the VI score, the colour transition point from red to blue is set at 1.2 as this is the visual inspection score cut for candidates in category A. Hence, areas in red shades represent a region where we can recover category A lens candidates and areas in blue shades represent a region where as visual inspectors we struggle to properly recognize or miss lens candidates. }
\label{fig:selfunc}
\end{figure}

The simulations in the test set used in stage-2 provide a broad, though not exhaustive, insight into our selection function. In Fig.~\ref{fig:selfunc} , we present each simulation alongside its corresponding visual inspection score, mapped within the parameter space of Einstein radii and the SNR of the lensed source in the \IE band. To better understand the relationship between these parameters and the visual inspection score, we used a Gaussian process regressor (GPR) from the \texttt{scikit-learn} library \citep{scikit-learn}. The GPR allows us to predict scores across the parameter range, therefore to understand the pattern in the data to model it and account for uncertainties. To do this we used a composite kernel consisting of a “ConstantKernel” which represents the overall scale of the parameter function, a “MaternKernel” which provides flexibility in modelling smooth variations and a “WhiteKernel” that accounts for noise in the data. We use the GPR to predict scores across the Einstein radii and the SNR range. This allows us to create contour levels that provide a broad approximation of the expected score for each lens based on the SNR and Einstein radius alone. With this we identify the regions of simulated lenses where we successfully classify lens candidates versus those where they are missed.

\begin{figure*}[htbp!]
\centering
\includegraphics[angle=0,width=1.0\hsize]{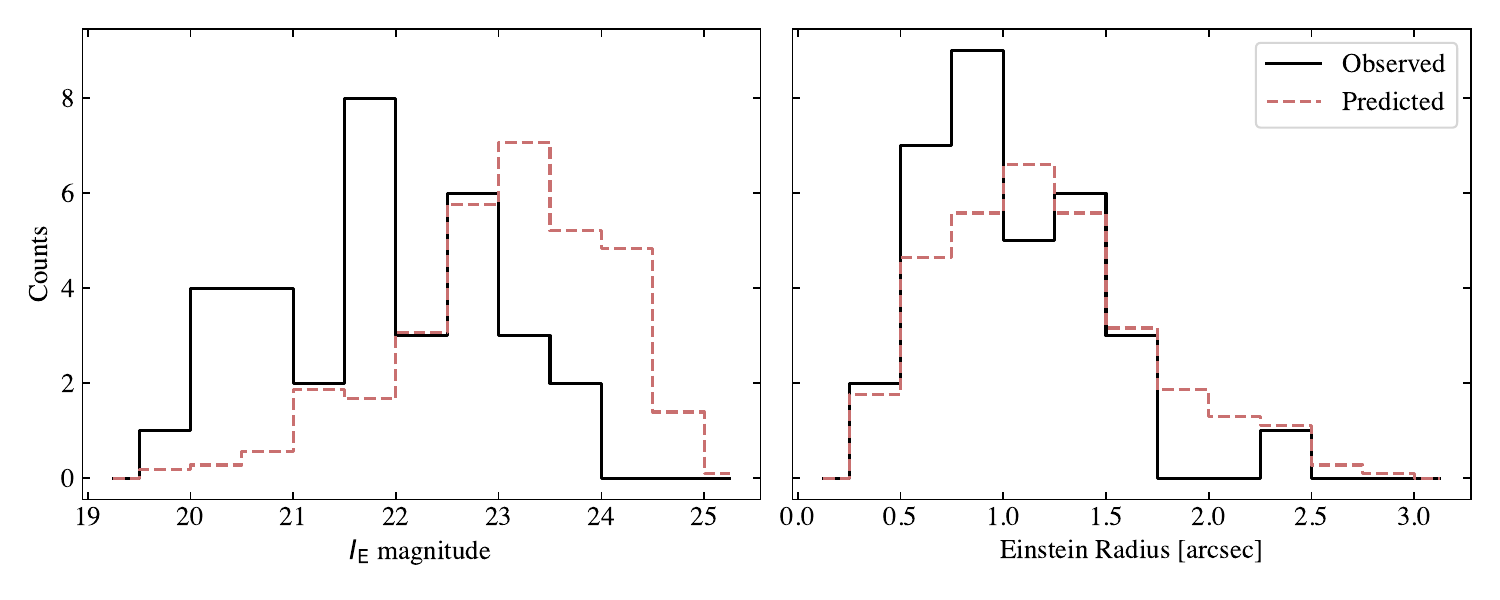}
\caption{\IE magnitude of the lensed source and Einstein radii distributions of lens candidates obtained after automatic lens modeling (black) and {\sc{LensPop}} predicted population, given the redshift and velocity dispersions of our initial sample (red).}
\label{fig:einsteinraddist}
\end{figure*}

From the simulations, we predict that most of our highly scored candidates will have a high SNR and large Einstein radii, while systems with low SNR and small Einstein radii are the most likely to be missed by visual inspectors. This prediction is confirmed when we analyze the model parameters of our lens candidates, although we have much sparser coverage. The Einstein radii distribution of our candidates peaks at 1\arcsecond\ (see Fig.~\ref{fig:einsteinraddist}), though  it is important to remember that we preselected high-velocity dispersion galaxies, making small Einstein radius configurations less probable.
Regarding SNR, the trend is clear: higher SNR correlates with higher visual inspection scores, and thus a greater probability of being recognized by visual inspectors. This is expected, as higher SNR ensures the lensing features are visible, but also highlights the limitations of human visual inspection. These results align with our expectations, as systems with low SNR or small Einstein radii pose significant challenges for human visual inspection \citep{Rojas+2022}. However, it is clear from Fig.~\ref{fig:einsteinraddist} that the sample does not perfectly match what was predicted by {\sc{LensPop}}: the Einstein radii are slightly smaller and the arcs are substantially brighter. The difference in Einstein radii is likely because we neglected uncertainties in the observed velocity dispersions. There are far more low-mass galaxies that could scatter up from below our $180\,\kms$ cut than go in the other direction. The brighter than expected VIS arc magnitudes hint that the definition of a discoverable lens and the source population in {\sc{LensPop}} are systematically incorrect. Though the fact that the total number of lenses discovered is comparable to the $\sim30$ predicted in Sect.~\ref{sc:forecast}, suggests that these effects somewhat cancel out.

\subsection{\label{sc:trainingset} A training set for machine learning}

In this section we will present some of the training samples used for the machine learning models and visual inspection projects runned in Q1, with a special emphasis on the improvements implemented in the simulation procedure.

Data driven simulations are a powerful set to train machine learning models and also test the performance of humans involved in visual inspections projects. To benefit, simulations need to be realistic enough, to teach the right properties to neural networks and to convince the human eye, hence for Q1 we worked on two main improvements compared with the dataset presented during the visual inspection project described here: better information matching \Euclid infrared bands for source magnitudes and utilization of the corresponding PSF. 

First, to properly match the magnitudes of the sources in the infrared bands, we used the COSMOS 2020 \citep{weaver2022} catalogue and followed the same procedure as before but this time using the VISTA $Y$, $J$, $H$-bands to match \Euclid's \YE, \JE, \HE  bands, resulting in a more realistic colour composite version of the simulations. 

Secondly, to transform the lensed source image into the \Euclid properties, instead of using a circular Gaussian to mimic the effect of the PSF we used the modelled PSF from the \Euclid pipeline for each cutout where we added a lensed source. Using this results in a lensed source that better matches the properties of the \Euclid image and prevents us creating unrealistic lensing sources that are too sharp or too smooth. 

A total of 2585 LRGs categorized during stage-1 had Q1 available data. We use this sample to perform our new Q1 simulations. Additionally to provide a larger training set we rotate each LRG image by 90 degrees and we pair it with a different source to produce a unique new simulations. This method has been successfully applied before by \cite{schuldt2021,Schuldt2023}. Using this method we quadruplicate the original set providing a final training set with about $10\,000$ examples. 

\begin{figure}[htbp!]
\centering
\includegraphics[angle=0,width=1.0\hsize]{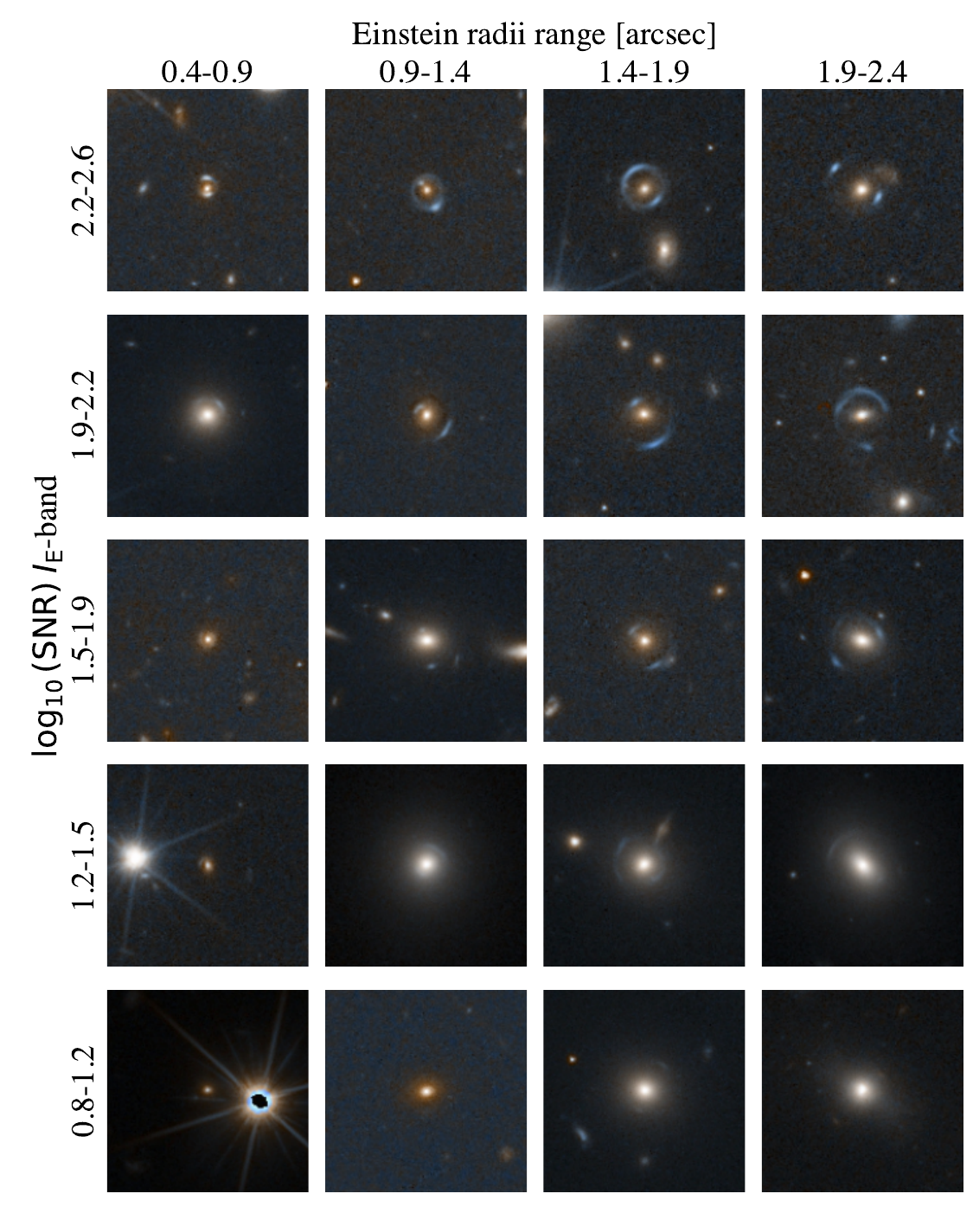}
\caption{Example of simulations following the new procedure. The 20 simulations are an example of a target in a different range of an Einstein radii and log$_{10}$(SNR) in \IE band. Each cutout has a size of $15\arcsec \times 15\arcsec$ and they are displayed using an MTF function using \IE and \YE bands.}
\label{fig:newsims}
\end{figure}

These new simulations as well as the catalogues of spirals, rings, mergers and other previously classified in this work were used to train different machine learning models \citep{Q1-SP048, Q1-SP053}. Simulations were also used to understand the selection function in the expert visual inspection and citizen science projects carried out in the Q1 lens finding project \citep{Q1-SP048,Q1-SP059}. In Fig.~\ref{fig:newsims} we present some examples of these simulations based on Q1 data, spanning a wider range than the ones created in stage-2.

\section{\label{sc:conclusions} Conclusion}

In this work we have shown that visual inspection of high-velocity dispersion galaxies is an efficient route to discovering large numbers of lenses, without the need for machine learning assistance. 

We inspected 11\,660 images and discovered 38 grade A and 40 grade B lenses. This is substantially more than were discovered in the untargeted inspection of \Euclid ERO data which found 3 grade A and 13 grade B in 12\,086 images \citep{AcevedoBarroso24}. Unlike an untargeted search, our approach will always miss low velocity dispersion lenses and lenses without spectroscopy, but it is substantially more efficient at finding lenses per human inspection.

We have 6 spectroscopically confirmed candidates. From Palomar Observatory we obtained source redshift for 5 lens systems. From DESI and SDSS we have redshifts for all the lens candidates, and additional redshift for one source from DESI.

The expected number of lenses in our sample was 32 (with substantial uncertainties), based on modifications of the forecasts of \citet{Collett2015}. It is not clear if we have found more candidates due to shot noise or because those forecasts neglect the lensing cross-section boost of group and cluster halos, or galaxy-galaxy lens rates are intrinsically higher than the \citet{Collett2015} model predicts. What is clear is that our sample is unlikely to be highly impure. Of the 21 grade A lenses for which the \Euclid automated lens modeller ran successfully \citep{Nightingale2021}, all of them are confirmed as lenses. Regarding grade B lenses 17 are confirmed as lenses with 6 excluded. The failure of the automatic modeller on the remaining candidates is not evidence that they are not lenses, as the modeller can fail on true lenses due to group scale halos or foreground light contamination.

Our approach cannot easily be scaled up to larger samples: DESI DR1 and \Euclid DR1 are not expected to show substantial overlap, and the visual inspection effort needed would be substantial even if we wait for the full datasets from both surveys. 

An equally important aspect of our approach was to expertly label a large sample of common false positives in machine learning based strong lens searches and to produce a sample of LRGs with known redshift and velocity dispersions that could be used to make a large sample of high-fidelity simulations of lenses by painting sources behind them. This result was a fantastic success, enabling us to produce a  sample of 10\,000 realistic simulated \Euclid images of lenses and 5366 false positives broken into spiral, ring galaxy, merger and other subclassification. 

On the metric of establishing a viable training set for machine learning, we have been hugely successful. Five teams trained using our sample \citep{Q1-SP053}, enabling citizen scientists and experts to efficiently discover 246 grade A and 254 grade B lenses \citep{Q1-SP048}. This galaxy-galaxy strong lensing discovery engine is ready to discover over 100\,000 strong lenses in the full \Euclid dataset. The visual inspection of spectroscopically selected lenses is the foundation stone of the \Euclid strong lensing revolution. 


%
%

\begin{acknowledgements}
\AckQone\\

\AckEC\\

K.R. acknowledge support from the Swiss National Science Foundation (SNSF) Grant Nr CRSII5\,198674. 

This work has received funding from the European Research Council (ERC) under the European Union's Horizon 2020 research and innovation programme (LensEra: grant agreement No 945536). TEC is funded by the Royal Society through a University Research Fellowship.

C.T. acknowledges the INAF grant 2022 LEMON.

Based on observations obtained at the Hale Telescope, Palomar Observatory, as part of a collaborative agreement between the Caltech Optical Observatories and the Jet Propulsion Laboratory.  We thank the following people who participated in the Palomar observing: Connor Auge, Indie Desiderio-Sloane, Jarred Gillette, Ollie Jackson, Grace Kallman, Michael Koss, Ai-Den Le, Alessandro Peca, Krysten Roldan, and Paul Shen.

This work used IRIS computing resources funded by the Science and Technology Facilities Council.

DESI construction and operations is managed by the Lawrence Berkeley National Laboratory. This research is supported by the U.S. Department of Energy, Office of Science, Office of High-Energy Physics, under Contract No. DE–AC02–05CH11231, and by the National Energy Research Scientific Computing Center, a DOE Office of Science User Facility under the same contract. Additional support for DESI is provided by the U.S. National Science Foundation, Division of Astronomical Sciences under Contract No. AST-0950945 to the NSF’s National Optical-Infrared Astronomy Research Laboratory; the Science and Technology Facilities Council of the United Kingdom; the Gordon and Betty Moore Foundation; the Heising-Simons Foundation; the French Alternative Energies and Atomic Energy Commission (CEA); the National Council of Science and Technology of Mexico (CONACYT); the Ministry of Science and Innovation of Spain, and by the DESI Member Institutions. The DESI collaboration is honored to be permitted to conduct astronomical research on Iolkam Du'ag (Kitt Peak), a mountain with particular significance to the Tohono O'odham Nation.

\end{acknowledgements}

%
%

\bibliography{my, Euclid, Q1}

\begin{thebibliography}{63}
\expandafter\ifx\csname natexlab\endcsname\relax\def\natexlab#1{#1}\fi

\bibitem[{{Acevedo Barroso} {et~al.}(2024){Acevedo Barroso}, {O'Riordan}, {Cl{\'e}ment}, {et~al.}}]{AcevedoBarroso24}
{Acevedo Barroso}, J.~A., {O'Riordan}, C.~M., {Cl{\'e}ment}, B., {et~al.} 2024, A\&A, submitted, arXiv:2408.06217

\bibitem[{{Aihara} {et~al.}(2018){Aihara}, {Arimoto}, {Armstrong}, {Arnouts}, {Bahcall}, {Bickerton}, {Bosch}, {Bundy}, {Capak}, {Chan}, {Chiba}, {Coupon}, {Egami}, {Enoki}, {Finet}, {Fujimori}, {Fujimoto}, {Furusawa}, {Furusawa}, {Goto}, {Goulding}, {Greco}, {Greene}, {Gunn}, {Hamana}, {Harikane}, {Hashimoto}, {Hattori}, {Hayashi}, {Hayashi}, {He{\l}miniak}, {Higuchi}, {Hikage}, {Ho}, {Hsieh}, {Huang}, {Huang}, {Ikeda}, {Imanishi}, {Inoue}, {Iwasawa}, {Iwata}, {Jaelani}, {Jian}, {Kamata}, {Karoji}, {Kashikawa}, {Katayama}, {Kawanomoto}, {Kayo}, {Koda}, {Koike}, {Kojima}, {Komiyama}, {Konno}, {Koshida}, {Koyama}, {Kusakabe}, {Leauthaud}, {Lee}, {Lin}, {Lin}, {Lupton}, {Mandelbaum}, {Matsuoka}, {Medezinski}, {Mineo}, {Miyama}, {Miyatake}, {Miyazaki}, {Momose}, {More}, {More}, {Moritani}, {Moriya}, {Morokuma}, {Mukae}, {Murata}, {Murayama}, {Nagao}, {Nakata}, {Niida}, {Niikura}, {Nishizawa}, {Obuchi}, {Oguri}, {Oishi}, {Okabe}, {Okamoto}, {Okura}, {Ono}, {Onodera}, {Onoue}, {Osato}, {Ouchi}, {Price}, {Pyo},
  {Sako}, {Sawicki}, {Shibuya}, {Shimasaku}, {Shimono}, {Shirasaki}, {Silverman}, {Simet}, {Speagle}, {Spergel}, {Strauss}, {Sugahara}, {Sugiyama}, {Suto}, {Suyu}, {Suzuki}, {Tait}, {Takada}, {Takata}, {Tamura}, {Tanaka}, {Tanaka}, {Tanaka}, {Tanaka}, {Terai}, {Terashima}, {Toba}, {Tominaga}, {Toshikawa}, {Turner}, {Uchida}, {Uchiyama}, {Umetsu}, {Uraguchi}, {Urata}, {Usuda}, {Utsumi}, {Wang}, {Wang}, {Wong}, {Yabe}, {Yamada}, {Yamanoi}, {Yasuda}, {Yeh}, {Yonehara}, \& {Yuma}}]{Aihara2018}
{Aihara}, H., {Arimoto}, N., {Armstrong}, R., {et~al.} 2018, \pasj, 70, S4

\bibitem[{{Almeida} {et~al.}(2023){Almeida}, {Anderson}, {Argudo-Fern{\'a}ndez}, {Badenes}, {Barger}, {Barrera-Ballesteros}, {Bender}, {Benitez}, {Besser}, {Bird}, {Bizyaev}, {Blanton}, {Bochanski}, {Bovy}, {Brandt}, {Brownstein}, {Buchner}, {Bulbul}, {Burchett}, {Cano D{\'\i}az}, {Carlberg}, {Casey}, {Chandra}, {Cherinka}, {Chiappini}, {Coker}, {Comparat}, {Conroy}, {Contardo}, {Cortes}, {Covey}, {Crane}, {Cunha}, {Dabbieri}, {Davidson}, {Davis}, {de Andrade Queiroz}, {De Lee}, {M{\'e}ndez Delgado}, {Demasi}, {Di Mille}, {Donor}, {Dow}, {Dwelly}, {Eracleous}, {Eriksen}, {Fan}, {Farr}, {Frederick}, {Fries}, {Frinchaboy}, {G{\"a}nsicke}, {Ge}, {Gonz{\'a}lez {\'A}vila}, {Grabowski}, {Grier}, {Guiglion}, {Gupta}, {Hall}, {Hawkins}, {Hayes}, {Hermes}, {Hern{\'a}ndez-Garc{\'\i}a}, {Hogg}, {Holtzman}, {Ibarra-Medel}, {Ji}, {Jofre}, {Johnson}, {Jones}, {Kinemuchi}, {Kluge}, {Koekemoer}, {Kollmeier}, {Kounkel}, {Krishnarao}, {Krumpe}, {Lacerna}, {Lago}, {Laporte}, {Liu}, {Liu}, {Liu}, {Lopes}, {Macktoobian},
  {Majewski}, {Malanushenko}, {Maoz}, {Masseron}, {Masters}, {Matijevic}, {McBride}, {Medan}, {Merloni}, {Morrison}, {Myers}, {M{\'e}sz{\'a}ros}, {Negrete}, {Nidever}, {Nitschelm}, {Oravetz}, {Oravetz}, {Pan}, {Peng}, {Pinsonneault}, {Pogge}, {Qiu}, {Ramirez}, {Rix}, {Fern{\'a}ndez Rosso}, {Runnoe}, {Salvato}, {Sanchez}, {Santana}, {Saydjari}, {Sayres}, {Schlaufman}, {Schneider}, {Schwope}, {Serna}, {Shen}, {Sobeck}, {Song}, {Souto}, {Spoo}, {Stassun}, {Steinmetz}, {Straumit}, {Stringfellow}, {S{\'a}nchez-Gallego}, {Taghizadeh-Popp}, {Tayar}, {Thakar}, {Tissera}, {Tkachenko}, {Hernandez Toledo}, {Trakhtenbrot}, {Fern{\'a}ndez-Trincado}, {Troup}, {Trump}, {Tuttle}, {Ulloa}, {Vazquez-Mata}, {Vera Alfaro}, {Villanova}, {Wachter}, {Weijmans}, {Wheeler}, {Wilson}, {Wojno}, {Wolf}, {Xue}, {Ybarra}, {Zari}, \& {Zasowski}}]{SDSS_DR18}
{Almeida}, A., {Anderson}, S.~F., {Argudo-Fern{\'a}ndez}, M., {et~al.} 2023, \apjs, 267, 44

\bibitem[{{Auger} {et~al.}(2009){Auger}, {Treu}, {Bolton}, {Gavazzi}, {Koopmans}, {Marshall}, {Bundy}, \& {Moustakas}}]{Auger2009}
{Auger}, M.~W., {Treu}, T., {Bolton}, A.~S., {et~al.} 2009, \apj, 705, 1099

\bibitem[{{Birrer} \& {Amara}(2018)}]{Birrer2018}
{Birrer}, S. \& {Amara}, A. 2018, Physics of the Dark Universe, 22, 189

\bibitem[{Birrer {et~al.}(2021)Birrer, Shajib, Gilman, Galan, Aalbers, Millon, Morgan, Pagano, Park, Teodori, Tessore, Ueland, de~Vyvere, Wagner-Carena, Wempe, Yang, Ding, Schmidt, Sluse, Zhang, \& Amara}]{Birrer2021}
Birrer, S., Shajib, A.~J., Gilman, D., {et~al.} 2021, Journal of Open Source Software, 6, 3283

\bibitem[{{Ca{\~n}ameras} {et~al.}(2020){Ca{\~n}ameras}, {Schuldt}, {Suyu}, {Taubenberger}, {Meinhardt}, {Leal-Taix{\'e}}, {Lemon}, {Rojas}, \& {Savary}}]{Canameras2020}
{Ca{\~n}ameras}, R., {Schuldt}, S., {Suyu}, S.~H., {et~al.} 2020, \aap, 644, A163

\bibitem[{{Cao} {et~al.}(2020){Cao}, {Li}, {Shu}, {Mao}, {Kneib}, \& {Gao}}]{cao2020}
{Cao}, X., {Li}, R., {Shu}, Y., {et~al.} 2020, \mnras, 499, 3610

\bibitem[{{Choi} {et~al.}(2007){Choi}, {Park}, \& {Vogeley}}]{Choi_park_vogeley2007}
{Choi}, Y.-Y., {Park}, C., \& {Vogeley}, M.~S. 2007, \apj, 658, 884

\bibitem[{{Collett}(2015)}]{Collett2015}
{Collett}, T.~E. 2015, \apj, 811, 20

\bibitem[{Collett {et~al.}(2018)Collett, Oldham, Smith, Auger, Westfall, Bacon, Nichol, Masters, Koyama, \& van~den Bosch}]{Collett_2018}
Collett, T.~E., Oldham, L.~J., Smith, R.~J., {et~al.} 2018, Science, 360, 1342

\bibitem[{{Connolly} {et~al.}(2010){Connolly}, {Peterson}, {Jernigan}, {Abel}, {Bankert}, {Chang}, {Claver}, {Gibson}, {Gilmore}, {Grace}, {Jones}, {Ivezic}, {Jee}, {Juric}, {Kahn}, {Krabbendam}, {Krughoff}, {Lorenz}, {Pizagno}, {Rasmussen}, {Todd}, {Tyson}, \& {Young}}]{Connoly2010}
{Connolly}, A.~J., {Peterson}, J., {Jernigan}, J.~G., {et~al.} 2010, in Society of Photo-Optical Instrumentation Engineers (SPIE) Conference Series, Vol. 7738, Modeling, Systems Engineering, and Project Management for Astronomy IV, ed. G.~Z. {Angeli} \& P.~{Dierickx}, 77381O

\bibitem[{{DESI Collaboration} {et~al.}(2024){DESI Collaboration}, {Adame}, {Aguilar}, {Ahlen}, {Alam}, {Aldering}, {Alexander}, {Alfarsy}, {Allende Prieto}, {Alvarez}, {Alves}, {Anand}, {Andrade-Oliveira}, {Armengaud}, {Asorey}, {Avila}, {Aviles}, {Bailey}, {Balaguera-Antol{\'\i}nez}, {Ballester}, {Baltay}, {Bault}, {Bautista}, {Behera}, {Beltran}, {BenZvi}, {Beraldo e Silva}, {Bermejo-Climent}, {Berti}, {Besuner}, {Beutler}, {Bianchi}, {Blake}, {Blum}, {Bolton}, {Brieden}, {Brodzeller}, {Brooks}, {Brown}, {Buckley-Geer}, {Burtin}, {Cabayol-Garcia}, {Cai}, {Canning}, {Cardiel-Sas}, {Carnero Rosell}, {Castander}, {Cervantes-Cota}, {Chabanier}, {Chaussidon}, {Chaves-Montero}, {Chen}, {Chen}, {Chuang}, {Claybaugh}, {Cole}, {Cooper}, {Cuceu}, {Davis}, {Dawson}, {de Belsunce}, {de la Cruz}, {de la Macorra}, {Della Costa}, {de Mattia}, {Demina}, {Demirbozan}, {DeRose}, {Dey}, {Dey}, {Dhungana}, {Ding}, {Ding}, {Doel}, {Doshi}, {Douglass}, {Edge}, {Eftekharzadeh}, {Eisenstein}, {Elliott}, {Ereza}, {Escoffier},
  {Fagrelius}, {Fan}, {Fanning}, {Fawcett}, {Ferraro}, {Flaugher}, {Font-Ribera}, {Forero-Romero}, {Forero-S{\'a}nchez}, {Frenk}, {G{\"a}nsicke}, {Garc{\'\i}a}, {Garc{\'\i}a-Bellido}, {Garcia-Quintero}, {Garrison}, {Gil-Mar{\'\i}n}, {Golden-Marx}, {Gontcho A Gontcho}, {Gonzalez-Morales}, {Gonzalez-Perez}, {Gordon}, {Graur}, {Green}, {Gruen}, {Guy}, {Hadzhiyska}, {Hahn}, {Han}, {Hanif}, {Herrera-Alcantar}, {Honscheid}, {Hou}, {Howlett}, {Huterer}, {Ir{\v{s}}i{\v{c}}}, {Ishak}, {Jacques}, {Jana}, {Jiang}, {Jimenez}, {Jing}, {Joudaki}, {Joyce}, {Jullo}, {Juneau}, {Kara{\c{c}}ayl{\i}}, {Karim}, {Kehoe}, {Kent}, {Khederlarian}, {Kim}, {Kirkby}, {Kisner}, {Kitaura}, {Kizhuprakkat}, {Kneib}, {Koposov}, {Kov{\'a}cs}, {Kremin}, {Krolewski}, {L'Huillier}, {Lahav}, {Lambert}, {Lamman}, {Lan}, {Landriau}, {Lang}, {Lange}, {Lasker}, {Leauthaud}, {Le Guillou}, {Levi}, {Li}, {Linder}, {Lyons}, {Magneville}, {Manera}, {Manser}, {Margala}, {Martini}, {McDonald}, {Medina}, {Medina-Varela}, {Meisner}, {Mena-Fern{\'a}ndez},
  {Meneses-Rizo}, {Mezcua}, {Miquel}, {Montero-Camacho}, {Moon}, {Moore}, {Moustakas}, {Mueller}, {Mundet}, {Mu{\~n}oz-Guti{\'e}rrez}, {Myers}, {Nadathur}, {Napolitano}, {Neveux}, {Newman}, {Nie}, {Nikutta}, {Niz}, {Norberg}, {Noriega}, {Paillas}, {Palanque-Delabrouille}, {Palmese}, {Pan}, {Parkinson}, {Penmetsa}, {Percival}, {P{\'e}rez-Fern{\'a}ndez}, {P{\'e}rez-R{\`a}fols}, {Pieri}, {Poppett}, {Porredon}, \& {Pothier}}]{DESI_EDR}
{DESI Collaboration}, {Adame}, A.~G., {Aguilar}, J., {et~al.} 2024, \aj, 168, 58

\bibitem[{{Euclid Collaboration: Aussel} {et~al.}(2025){Euclid Collaboration: Aussel}, {Tereno}, {Schirmer}, {et~al.}}]{Q1-TP001}
{Euclid Collaboration: Aussel}, H., {Tereno}, I., {Schirmer}, M., {et~al.} 2025, \aap, submitted

\bibitem[{{Euclid Collaboration: Cropper} {et~al.}(2024){Euclid Collaboration: Cropper}, {Al Bahlawan}, {Amiaux}, {et~al.}}]{EuclidSkyVIS}
{Euclid Collaboration: Cropper}, M., {Al Bahlawan}, A., {Amiaux}, J., {et~al.} 2024, \aap, accepted, arXiv:2405.13492

\bibitem[{{Euclid Collaboration: Holloway} {et~al.}(2025){Euclid Collaboration: Holloway}, {Verma}, {Walmsley}, {et~al.}}]{Q1-SP059}
{Euclid Collaboration: Holloway}, P., {Verma}, A., {Walmsley}, M., {et~al.} 2025, \aap, submitted

\bibitem[{{Euclid Collaboration: Jahnke} {et~al.}(2024){Euclid Collaboration: Jahnke}, {Gillard}, {Schirmer}, {et~al.}}]{EuclidSkyNISP}
{Euclid Collaboration: Jahnke}, K., {Gillard}, W., {Schirmer}, M., {et~al.} 2024, \aap, accepted, arXiv:2405.13493

\bibitem[{{Euclid Collaboration: Li} {et~al.}(2025){Euclid Collaboration: Li}, {Collett}, {Walmsley}, {et~al.}}]{Q1-SP054}
{Euclid Collaboration: Li}, T., {Collett}, T., {Walmsley}, M., {et~al.} 2025, \aap, submitted

\bibitem[{{Euclid Collaboration: Lines} {et~al.}(2025){Euclid Collaboration: Lines}, {Collett}, {Walmsley}, {et~al.}}]{Q1-SP053}
{Euclid Collaboration: Lines}, N. E.~P., {Collett}, T.~E., {Walmsley}, M., {et~al.} 2025, \aap, submitted

\bibitem[{{Euclid Collaboration: Mellier} {et~al.}(2024){Euclid Collaboration: Mellier}, {Abdurro'uf}, {Acevedo~Barroso}, {et~al.}}]{EuclidSkyOverview}
{Euclid Collaboration: Mellier}, Y., {Abdurro'uf}, {Acevedo~Barroso}, J., {et~al.} 2024, \aap, accepted, arXiv:2405.13491

\bibitem[{{Euclid Collaboration: Walmsley} {et~al.}(2025){Euclid Collaboration: Walmsley}, {Holloway}, {Lines}, {et~al.}}]{Q1-SP048}
{Euclid Collaboration: Walmsley}, M., {Holloway}, P., {Lines}, N., {et~al.} 2025, \aap, submitted

\bibitem[{{Euclid Quick Release Q1}(2025)}]{Q1cite}
{Euclid Quick Release Q1}. 2025, \url{https://doi.org/10.57780/esa-2853f3b}

\bibitem[{Faure {et~al.}(2008)Faure, Kneib, Covone, Tasca, Leauthaud, Capak, Jahnke, Smolcic, de~la Torre, Ellis, Finoguenov, Koekemoer, Fevre, Massey, Mellier, Refregier, Rhodes, Scoville, Schinnerer, Taylor, Waerbeke, \& Walcher}]{Faure_2008}
Faure, C., Kneib, J.-P., Covone, G., {et~al.} 2008, ApJS, 176, 19

\bibitem[{Garvin {et~al.}(2022)Garvin, Kruk, Cornen, Bhatawdekar, Cañameras, \& Merín}]{Garvin_2022}
Garvin, E.~O., Kruk, S., Cornen, C., {et~al.} 2022, \aap, 667, A141

\bibitem[{{Goobar} {et~al.}(2017){Goobar}, {Amanullah}, {Kulkarni}, {Nugent}, {Johansson}, {Steidel}, {Law}, {M{\"o}rtsell}, {Quimby}, {Blagorodnova}, {Brandeker}, {Cao}, {Cooray}, {Ferretti}, {Fremling}, {Hangard}, {Kasliwal}, {Kupfer}, {Lunnan}, {Masci}, {Miller}, {Nayyeri}, {Neill}, {Ofek}, {Papadogiannakis}, {Petrushevska}, {Ravi}, {Sollerman}, {Sullivan}, {Taddia}, {Walters}, {Wilson}, {Yan}, \& {Yaron}}]{Goobar2017}
{Goobar}, A., {Amanullah}, R., {Kulkarni}, S.~R., {et~al.} 2017, Science, 356, 291

\bibitem[{{He} {et~al.}(2024){He}, {Nightingale}, {Amvrosiadis}, {Robertson}, {Cole}, {Frenk}, {Massey}, {Li}, {Cao}, {Lange}, \& {Fran{\c{c}}a}}]{He2024}
{He}, Q., {Nightingale}, J.~W., {Amvrosiadis}, A., {et~al.} 2024, \mnras, 532, 2441

\bibitem[{Ilbert {et~al.}(2008)Ilbert, Capak, Salvato, Aussel, McCracken, Sanders, Scoville, Kartaltepe, Arnouts, Floc'h, Mobasher, Taniguchi, Lamareille, Leauthaud, Sasaki, Thompson, Zamojski, Zamorani, Bardelli, Bolzonella, Bongiorno, Brusa, Caputi, Carollo, Contini, Cook, Coppa, Cucciati, de~la Torre, de~Ravel, Franzetti, Garilli, Hasinger, Iovino, Kampczyk, Kneib, Knobel, Kovac, Le~Borgne, Le~Brun, Fèvre, Lilly, Looper, Maier, Mainieri, Mellier, Mignoli, Murayama, Pellò, Peng, Pérez-Montero, Renzini, Ricciardelli, Schiminovich, Scodeggio, Shioya, Silverman, Surace, Tanaka, Tasca, Tresse, Vergani, \& Zucca}]{Ilbert_2009}
Ilbert, O., Capak, P., Salvato, M., {et~al.} 2008, ApJ, 690, 1236

\bibitem[{{Jackson}(2008)}]{Jackson2008}
{Jackson}, N. 2008, \mnras, 389, 1311

\bibitem[{{Jacobs} {et~al.}(2019){Jacobs}, {Collett}, {Glazebrook}, {Buckley-Geer}, {Diehl}, {Lin}, {McCarthy}, {Qin}, {Odden}, {Caso Escudero}, {Dial}, {Yung}, {Gaitsch}, {Pellico}, {Lindgren}, {Abbott}, {Annis}, {Avila}, {Brooks}, {Burke}, {Carnero Rosell}, {Carrasco Kind}, {Carretero}, {da Costa}, {De Vicente}, {Fosalba}, {Frieman}, {Garc{\'\i}a-Bellido}, {Gaztanaga}, {Goldstein}, {Gruen}, {Gruendl}, {Gschwend}, {Hollowood}, {Honscheid}, {Hoyle}, {James}, {Krause}, {Kuropatkin}, {Lahav}, {Lima}, {Maia}, {Marshall}, {Miquel}, {Plazas}, {Roodman}, {Sanchez}, {Scarpine}, {Serrano}, {Sevilla-Noarbe}, {Smith}, {Sobreira}, {Suchyta}, {Swanson}, {Tarle}, {Vikram}, {Walker}, {Zhang}, \& {DES Collaboration}}]{Jacobs2019}
{Jacobs}, C., {Collett}, T., {Glazebrook}, K., {et~al.} 2019, \apjs, 243, 17

\bibitem[{{Jacobs} {et~al.}(2017){Jacobs}, {Glazebrook}, {Collett}, {More}, \& {McCarthy}}]{Jacobs17}
{Jacobs}, C., {Glazebrook}, K., {Collett}, T., {More}, A., \& {McCarthy}, C. 2017, \mnras, 471, 167

\bibitem[{{Kelly} {et~al.}(2018){Kelly}, {Diego}, {Rodney}, {Kaiser}, {Broadhurst}, {Zitrin}, {Treu}, {P{\'e}rez-Gonz{\'a}lez}, {Morishita}, {Jauzac}, {Selsing}, {Oguri}, {Pueyo}, {Ross}, {Filippenko}, {Smith}, {Hjorth}, {Cenko}, {Wang}, {Howell}, {Richard}, {Frye}, {Jha}, {Foley}, {Norman}, {Bradac}, {Zheng}, {Brammer}, {Benito}, {Cava}, {Christensen}, {de Mink}, {Graur}, {Grillo}, {Kawamata}, {Kneib}, {Matheson}, {McCully}, {Nonino}, {P{\'e}rez-Fournon}, {Riess}, {Rosati}, {Schmidt}, {Sharon}, \& {Weiner}}]{Kelly2018}
{Kelly}, P.~L., {Diego}, J.~M., {Rodney}, S., {et~al.} 2018, Nature Astronomy, 2, 334

\bibitem[{{Kelly} {et~al.}(2015){Kelly}, {Rodney}, {Treu}, {Foley}, {Brammer}, {Schmidt}, {Zitrin}, {Sonnenfeld}, {Strolger}, {Graur}, {Filippenko}, {Jha}, {Riess}, {Bradac}, {Weiner}, {Scolnic}, {Malkan}, {von der Linden}, {Trenti}, {Hjorth}, {Gavazzi}, {Fontana}, {Merten}, {McCully}, {Jones}, {Postman}, {Dressler}, {Patel}, {Cenko}, {Graham}, \& {Tucker}}]{Kelly2015}
{Kelly}, P.~L., {Rodney}, S.~A., {Treu}, T., {et~al.} 2015, Science, 347, 1123

\bibitem[{{Koekemoer} {et~al.}(2007){Koekemoer}, {Aussel}, {Calzetti}, {Capak}, {Giavalisco}, {Kneib}, {Leauthaud}, {Le F{\`e}vre}, {McCracken}, {Massey}, {Mobasher}, {Rhodes}, {Scoville}, \& {Shopbell}}]{Koekemoer2007}
{Koekemoer}, A.~M., {Aussel}, H., {Calzetti}, D., {et~al.} 2007, \apjs, 172, 196

\bibitem[{{Kollmeier} {et~al.}(2019){Kollmeier}, {Anderson}, {Blanc}, {Blanton}, {Covey}, {Crane}, {Drory}, {Frinchaboy}, {Froning}, {Johnson}, {Kneib}, {Kreckel}, {Merloni}, {Pellegrini}, {Pogge}, {Ramirez}, {Rix}, {Sayres}, {S{\'a}nchez-Gallego}, {Shen}, {Tkachenko}, {Trump}, {Tuttle}, {Weijmans}, {Zasowski}, {Barbuy}, {Beaton}, {Bergemann}, {Bochanski}, {Brandt}, {Casey}, {Cherinka}, {Eracleous}, {Fan}, {Garc{\'\i}a}, {Green}, {Hekker}, {Lane}, {Longa-Pe{\~n}a}, {Mathur}, {Meza}, {Minchev}, {Myers}, {Nidever}, {Nitschelm}, {O'Connell}, {Price-Whelan}, {Raddick}, {Rossi}, {Sankrit}, {Simon}, {Stutz}, {Ting}, {Trakhtenbrot}, {Weaver}, {Willmer}, \& {Weinberg}}]{SDSSV_2019}
{Kollmeier}, J., {Anderson}, S.~F., {Blanc}, G.~A., {et~al.} 2019, in Bulletin of the American Astronomical Society, Vol.~51, 274

\bibitem[{{Leauthaud} {et~al.}(2007){Leauthaud}, {Massey}, {Kneib}, {Rhodes}, {Johnston}, {Capak}, {Heymans}, {Ellis}, {Koekemoer}, {Le F{\`e}vre}, {Mellier}, {R{\'e}fr{\'e}gier}, {Robin}, {Scoville}, {Tasca}, {Taylor}, \& {Van Waerbeke}}]{Leauthaud2007}
{Leauthaud}, A., {Massey}, R., {Kneib}, J.-P., {et~al.} 2007, \apjs, 172, 219

\bibitem[{{Li} {et~al.}(2020){Li}, {Napolitano}, {Tortora}, {Spiniello}, {Koopmans}, {Huang}, {Roy}, {Vernardos}, {Chatterjee}, {Giblin}, {Getman}, {Radovich}, {Covone}, \& {Kuijken}}]{Li2020}
{Li}, R., {Napolitano}, N.~R., {Tortora}, C., {et~al.} 2020, \apj, 899, 30

\bibitem[{Marshall {et~al.}(2015)Marshall, Verma, More, Davis, More, Kapadia, Parrish, Snyder, Wilcox, Baeten, Macmillan, Cornen, Baumer, Simpson, Lintott, Miller, Paget, Simpson, Smith, Küng, Saha, \& Collett}]{SpaceWarpsI}
Marshall, P.~J., Verma, A., More, A., {et~al.} 2015, MNRAS, 455, 1171

\bibitem[{Meena {et~al.}(2023)Meena, Zitrin, Jiménez-Teja, Zackrisson, Chen, Coe, Diego, Dimauro, Furtak, Kelly, Oguri, Welch, Abdurro’uf, Andrade-Santos, Adamo, Bhatawdekar, Bradač, Bradley, Broadhurst, Conselice, Dayal, Donahue, Frye, Fujimoto, Hsiao, Kokorev, Mahler, Vanzella, \& Windhorst}]{Meena_2023}
Meena, A.~K., Zitrin, A., Jiménez-Teja, Y., {et~al.} 2023, ApJL, 944, L6

\bibitem[{{More} {et~al.}(2012){More}, {Cabanac}, {More}, {Alard}, {Limousin}, {Kneib}, {Gavazzi}, \& {Motta}}]{More2012}
{More}, A., {Cabanac}, R., {More}, S., {et~al.} 2012, \apj, 749, 38

\bibitem[{{More} {et~al.}(2016){More}, {Verma}, {Marshall}, {More}, {Baeten}, {Wilcox}, {Macmillan}, {Cornen}, {Kapadia}, {Parrish}, {Snyder}, {Davis}, {Gavazzi}, {Lintott}, {Simpson}, {Miller}, {Smith}, {Paget}, {Saha}, {K{\"u}ng}, \& {Collett}}]{SpaceWarpsII}
{More}, A., {Verma}, A., {Marshall}, P.~J., {et~al.} 2016, \mnras, 455, 1191

\bibitem[{Nelder \& Mead(1965)}]{nelder_mead_1965}
Nelder, J.~A. \& Mead, R. 1965, The Computer Journal, 7, 308

\bibitem[{{Nightingale} {et~al.}(2021){Nightingale}, {Hayes}, {Kelly}, {Amvrosiadis}, {Etherington}, {He}, {Li}, {Cao}, {Frawley}, {Cole}, {Enia}, {Frenk}, {Harvey}, {Li}, {Massey}, {Negrello}, \& {Robertson}}]{Nightingale2021}
{Nightingale}, J., {Hayes}, R., {Kelly}, A., {et~al.} 2021, The Journal of Open Source Software, 6, 2825

\bibitem[{{Nightingale} {et~al.}(2024){Nightingale}, {Massey}, {Kegerreis}, \& {Hayes}}]{Nightingale2024}
{Nightingale}, J., {Massey}, R., {Kegerreis}, J., \& {Hayes}, R. 2024, The Journal of Open Source Software, 9, 4904

\bibitem[{{Oke} \& {Gunn}(1982)}]{Oke1982}
{Oke}, J.~B. \& {Gunn}, J.~E. 1982, \pasp, 94, 586

\bibitem[{{Pawase} {et~al.}(2014){Pawase}, {Courbin}, {Faure}, {Kokotanekova}, \& {Meylan}}]{Pawase2014}
{Pawase}, R.~S., {Courbin}, F., {Faure}, C., {Kokotanekova}, R., \& {Meylan}, G. 2014, \mnras, 439, 3392

\bibitem[{Pedregosa {et~al.}(2011)Pedregosa, Varoquaux, Gramfort, Michel, Thirion, Grisel, Blondel, Prettenhofer, Weiss, Dubourg, Vanderplas, Passos, Cournapeau, Brucher, Perrot, \& Duchesnay}]{scikit-learn}
Pedregosa, F., Varoquaux, G., Gramfort, A., {et~al.} 2011, Journal of Machine Learning Research, 12, 2825

\bibitem[{{Petrillo} {et~al.}(2019){Petrillo}, {Tortora}, {Vernardos}, {Koopmans}, {Verdoes Kleijn}, {Bilicki}, {Napolitano}, {Chatterjee}, {Covone}, {Dvornik}, {Erben}, {Getman}, {Giblin}, {Heymans}, {de Jong}, {Kuijken}, {Schneider}, {Shan}, {Spiniello}, \& {Wright}}]{Petrillo2019}
{Petrillo}, C.~E., {Tortora}, C., {Vernardos}, G., {et~al.} 2019, \mnras, 484, 3879

\bibitem[{Pierel {et~al.}(2024)Pierel, Newman, Dhawan, Gu, Joshi, Li, Schuldt, Strolger, Suyu, Caminha, Cohen, Diego, DŚilva, Ertl, Frye, Granata, Grillo, Koekemoer, Li, Robotham, Summers, Treu, Windhorst, Zitrin, Agarwal, Agrawal, Arendse, Belli, Burns, Cañameras, Chakrabarti, Chen, Collett, Coulter, Ellis, Engesser, Foo, Fox, Gall, Garuda, Gezari, Gomez, Glazebrook, Hjorth, Huang, Jha, Kamieneski, Kelly, Larison, Moustakas, Pascale, Pérez-Fournon, Petrushevska, Poidevin, Rest, Shahbandeh, Shajib, Siebert, Storfer, Talbot, Wang, Wevers, \& Zenati}]{Pierel_2024}
Pierel, J. D.~R., Newman, A.~B., Dhawan, S., {et~al.} 2024, ApJL, 967, L37

\bibitem[{{Pourrahmani} {et~al.}(2018){Pourrahmani}, {Nayyeri}, \& {Cooray}}]{Pourrahmani2018}
{Pourrahmani}, M., {Nayyeri}, H., \& {Cooray}, A. 2018, \apj, 856, 68

\bibitem[{{Rojas} {et~al.}(2022){Rojas}, {Savary}, {Cl{\'e}ment}, {Maus}, {Courbin}, {Lemon}, {Chan}, {Vernardos}, {Joseph}, {Ca{\~n}ameras}, \& {Galan}}]{Rojas+2022}
{Rojas}, K., {Savary}, E., {Cl{\'e}ment}, B., {et~al.} 2022, \aap, 668, A73

\bibitem[{{Savary} {et~al.}(2022){Savary}, {Rojas}, {Maus}, {Cl{\'e}ment}, {Courbin}, {Gavazzi}, {Chan}, {Lemon}, {Vernardos}, {Ca{\~n}ameras}, {Schuldt}, {Suyu}, {Cuillandre}, {Fabbro}, {Gwyn}, {Hudson}, {Kilbinger}, {Scott}, \& {Stone}}]{Savary+22}
{Savary}, E., {Rojas}, K., {Maus}, M., {et~al.} 2022, \aap, 666, A1

\bibitem[{{Schuldt} {et~al.}(2023){Schuldt}, {Suyu}, {Ca{\~n}ameras}, {Shu}, {Taubenberger}, {Ertl}, \& {Halkola}}]{Schuldt2023}
{Schuldt}, S., {Suyu}, S.~H., {Ca{\~n}ameras}, R., {et~al.} 2023, \aap, 673, A33

\bibitem[{{Schuldt} {et~al.}(2021){Schuldt}, {Suyu}, {Meinhardt}, {Leal-Taix{\'e}}, {Ca{\~n}ameras}, {Taubenberger}, \& {Halkola}}]{schuldt2021}
{Schuldt}, S., {Suyu}, S.~H., {Meinhardt}, T., {et~al.} 2021, \aap, 646, A126

\bibitem[{{Scoville} {et~al.}(2007){Scoville}, {Abraham}, {Aussel}, {Barnes}, {Benson}, {Blain}, {Calzetti}, {Comastri}, {Capak}, {Carilli}, {Carlstrom}, {Carollo}, {Colbert}, {Daddi}, {Ellis}, {Elvis}, {Ewald}, {Fall}, {Franceschini}, {Giavalisco}, {Green}, {Griffiths}, {Guzzo}, {Hasinger}, {Impey}, {Kneib}, {Koda}, {Koekemoer}, {Lefevre}, {Lilly}, {Liu}, {McCracken}, {Massey}, {Mellier}, {Miyazaki}, {Mobasher}, {Mould}, {Norman}, {Refregier}, {Renzini}, {Rhodes}, {Rich}, {Sanders}, {Schiminovich}, {Schinnerer}, {Scodeggio}, {Sheth}, {Shopbell}, {Taniguchi}, {Tyson}, {Urry}, {Van Waerbeke}, {Vettolani}, {White}, \& {Yan}}]{Scoville2007}
{Scoville}, N., {Abraham}, R.~G., {Aussel}, H., {et~al.} 2007, \apjs, 172, 38

\bibitem[{{Shajib} {et~al.}(2020){Shajib}, {Treu}, {Birrer}, \& {Sonnenfeld}}]{Shajib2020}
{Shajib}, A.~J., {Treu}, T., {Birrer}, S., \& {Sonnenfeld}, A. 2020, arXiv e-prints, arXiv:2008.11724

\bibitem[{{Sonnenfeld}(2024)}]{Sonnenfeld2024}
{Sonnenfeld}, A. 2024, \aap, 690, A325

\bibitem[{{Tessore} \& {Metcalf}(2015)}]{Tessore2015}
{Tessore}, N. \& {Metcalf}, R.~B. 2015, \aap, 580, A79

\bibitem[{{Treu} \& {Koopmans}(2004)}]{TreuKoopmans2004}
{Treu}, T. \& {Koopmans}, L. V.~E. 2004, \apj, 611, 739

\bibitem[{{Treu} {et~al.}(2022){Treu}, {Suyu}, \& {Marshall}}]{Treu2022}
{Treu}, T., {Suyu}, S.~H., \& {Marshall}, P.~J. 2022, \aapr, 30, 8

\bibitem[{{Walsh} {et~al.}(1979){Walsh}, {Carswell}, \& {Weymann}}]{Walsh1979}
{Walsh}, D., {Carswell}, R.~F., \& {Weymann}, R.~J. 1979, \nat, 279, 381

\bibitem[{{Weaver} {et~al.}(2022){Weaver}, {Kauffmann}, {Ilbert}, {McCracken}, {Moneti}, {Toft}, {Brammer}, {Shuntov}, {Davidzon}, {Hsieh}, {Laigle}, {Anastasiou}, {Jespersen}, {Vinther}, {Capak}, {Casey}, {McPartland}, {Milvang-Jensen}, {Mobasher}, {Sanders}, {Zalesky}, {Arnouts}, {Aussel}, {Dunlop}, {Faisst}, {Franx}, {Furtak}, {Fynbo}, {Gould}, {Greve}, {Gwyn}, {Kartaltepe}, {Kashino}, {Koekemoer}, {Kokorev}, {Le F{\`e}vre}, {Lilly}, {Masters}, {Magdis}, {Mehta}, {Peng}, {Riechers}, {Salvato}, {Sawicki}, {Scarlata}, {Scoville}, {Shirley}, {Silverman}, {Sneppen}, {Smolc̆i{\'c}}, {Steinhardt}, {Stern}, {Tanaka}, {Taniguchi}, {Teplitz}, {Vaccari}, {Wang}, \& {Zamorani}}]{weaver2022}
{Weaver}, J.~R., {Kauffmann}, O.~B., {Ilbert}, O., {et~al.} 2022, \apjs, 258, 11

\bibitem[{{Welch} {et~al.}(2022){Welch}, {Coe}, {Diego}, {Zitrin}, {Zackrisson}, {Dimauro}, {Jim{\'e}nez-Teja}, {Kelly}, {Mahler}, {Oguri}, {Timmes}, {Windhorst}, {Florian}, {de Mink}, {Avila}, {Anderson}, {Bradley}, {Sharon}, {Vikaeus}, {McCandliss}, {Brada{\v{c}}}, {Rigby}, {Frye}, {Toft}, {Strait}, {Trenti}, {Sharma}, {Andrade-Santos}, \& {Broadhurst}}]{Welch2022}
{Welch}, B., {Coe}, D., {Diego}, J.~M., {et~al.} 2022, \nat, 603, 815

\bibitem[{Wells {et~al.}(2024)Wells, Fassnacht, Birrer, \& Williams}]{Wells_2024}
Wells, P.~R., Fassnacht, C.~D., Birrer, S., \& Williams, D. 2024, \aap, 689, A87

\end{thebibliography}

%

\begin{appendix}
  \onecolumn 

\end{appendix}

\end{document}